\documentclass[twocolumn,amsmath,amssymb]{revtex4-2}
\usepackage{graphicx}
\usepackage{dcolumn}
\usepackage{color}
\usepackage{bm}
\usepackage{amsmath,amssymb,graphicx}
\usepackage{epsfig}
\usepackage{enumerate}
\usepackage{array}
\usepackage{multirow}
\begin {document}
\title{Microscopic analysis of relaxation behavior in nonlinear optical conductivity of graphene}
\author{Bristi Ghosh$^1$, Sushanta Dattagupta $^2$, Malay Bandyopadhyay $^1$ }
\affiliation{1.School of Basic Sciences, Indian Institute of Technology Bhubaneswar, Argul, Jatni, Khurda, Odisha 752050, India.\\
2. National Institute of Technology, Mahatma Gandhi Road, Durgapur, West Bengal, 713209, India}
\date{\today}
\begin{abstract}
We present here a general formulation for the interband dynamical optical conductivity in the nonlinear regime of graphene in the presence of a quantum bath comprising phonons and electrons. Our main focus is the relaxation behavior of the quantum solid of graphene perturbed by an oscillatory electric field. Considering the optical range of the frequency and a considerable amount of the amplitude of the field, one can observe a nonlinear response by formulating a quantum master equation of the density operator associated with the Hamiltonian encapsulated in the form of a spin-Boson model of dissipative quantum statistical mechanics. Mapping the valence and conduction states as the eigenstates of the Pauli spin operators and utilizing the rotating wave approximation to omit off-resonant terms, one can solve the rate equation for the mean population of the conduction and valence states and the mixing matrix elements between them.  Our results reveal the nonlinear steady-state regime's population inversion and interband coherence. It is characterized by a single dimensionless parameter that is directly proportional to the incident field strength and inversely proportional to the optical frequency. Our method is also capable of calculating the nonlinear interband optical conductivity of doped and gapped graphene at finite temperatures. The effects of different bath spectra for phonons and electrons are examined in detail.
 Although our general formulation can address a variety of nonequilibrium response of the two-band system, it also facilitates a connection with phenomenological modeling of nonlinear optical conductivity.
\end{abstract}
\maketitle
\section{Introduction}
Graphene, a two-dimensional sheet of graphite, is a wonder laboratory of modern solid state physics that is endowed with remarkable physical properties and potential device applications [1,2]. It is a nano material in which much of fundamental predictions of relativistic quantum mechanics, such as the  Dirac equation, Weyl and Majorana electrons, geometric and topological phases, spintronics, etc., can be experimentally tested. Graphene, a true  two dimensional electronic material is not only a gold mine for their myriad technological and device applications, but also a repository for testing theoretical concepts of great contemporary interest [1]. One can mention some of these exceptional ideas such as chemistry of hybridized carbon orbitals [3], ultra high
mobility [4,5], spin-orbit interaction [6], Andreev reflection and Klein tunneling [7,8], magneto-resistance and
weak localization [9,10], quantum Hall effect [11,12], spintronics [13], and so on. Much of these extraordinary properties of graphene emanate from the fact that the electrons of graphene behave as Dirac Fermions in the low energy physics sector which exhibit typical linear band structure ($E_{p}=\pm v_{F}|p|$, where $v_F$ is the Fermi velocity) at $K$ and $K'$ points and follow Dirac physics. Although these electrons move with much smaller Fermi velocity compared to the speed of light, but their dynamics is governed by Dirac equation. Hence, this fascinating 2D-material becomes a testing-bed for the realization of relativistic quantum mechanics
in a non-relativistic setup of solid state physics [3]. In the presence of a strong electromagnetic field, these massless carriers display fascinating linear and nonlinear optical properties such as constant absorption coefficient over a broad spectrum [14], higher-harmonic generation [15], four-wave mixing [16], and self-phase modulation [17], just to name a few.\\
\indent
Given these extraordinary phenomena in graphene, we focus on the nonlinear response of frequency-dependent dynamic conductivity [18,19,20]. From the beginning, much attention has been devoted to the linear response of graphene to the applied electric field in the so-called Kubo regime. However, additional insights can be gained by transiting to the nonlinear response, especially when the applied electric field is dependent on a monochromatic frequency $\omega$.The nonlinear response of graphene is an exceptional tool for investigating intrinsic material properties that are hidden in the Kubo regime, such as material symmetry, selection rules, electron spin, and spin-spin relaxations mechanism [21]. The optical conductivity in the Kubo regime is clearly defined by the universal value $\sigma_0=\frac{e^{2}}{4\hbar}$, however, here we are more concerned with determining the frequency dependent conductivity in the nonlinear domain in the optical range ($10^{11}-10^{16}$ Hz). Furthermore, we take a close view into the relaxation behavior of the carriers associated with energy transfer between the applied field and elementary excitations characterizing the surrounding heat bath. Thus, we couch the problem in the contemporary field of nonequilibrium statistical mechanics of dissipative quantum systems.\\
\indent
The nonlinear optical response in the background of dissipative features had been looked into earlier in terms of rate theories familiar in quantum optics [18,19]. The study of Rabi oscillations, rotating wave approximation naturally feature into such theories. Two distinct relaxation attributes also have merited attention: spin-lattice relaxation ($\gamma_p$) and spin-spin relaxation ($\gamma_e$), common to magnetic resonance phenomena [22]. It is pertinent to point out that the ‘spin’ here refers to a pseudo-spin that captures the valence and the conduction band near the Dirac point in the reciprocal space. Spin-lattice relaxation is accompanied by inter-band transitions while spin-spin relaxations arise from intra-band transitions. An important quantity which clearly captures the nonlinear and frequency-dependent features is the so-called Mischenko parameter defined by $\eta=\frac{ev_FE_0}{\hbar\omega\sqrt{\gamma_e\gamma_p}}$, where $E_0$ and $\omega$ are the amplitude and frequency of the externally applied oscillatory field [18]. The latter demarcates the boundaries between linear and nonlinear domains. While the rate equation approach does provide significant insights into the phenomena at hand, such an approach has limitations in that the external bath is viewed as a ‘black-box’ and no attempt is made to give a microscopic assessment of the underlying relaxation rates. Our aim is to fill-in this gap and put forward a general master equation method for the underlying density operator of the system that goes beyond the rate theories. This was attempted by one of us [20] wherein a careful delineation was made between the non-Markov and Markov regions of relaxation and contact was established with the Markovian regime in which the rate theories are valid. We now go beyond [20] and analyze in detail the underlying spin-lattice and spin-spin relaxation rates characterized by the parameters of both the system and the bath, including the temperature
(T). Needless to say such temperature variations of the rates, that can be accessed experimentally, are beyond the realm of rate equation methods. Our numerical calculations enable us to quantify the crossover between the transient, non-Markovian response to non-transient, Markovian response as a function of a timescale governed by the cutoff frequency of bath excitations. Additionally, we demonstrate the temperature variation of both the spin-lattice and the spin-spin relaxation rates. We also go into the case of gapped graphene that brings in a new energy parameter ($\Delta$) that couples to the spin component transverse to the graphene layer. Results for the conductivity related to inter-band transitions are presented for both pristine (gapless) and gapped graphene. A novel switching behavior in the low temperature optical conductivity for gapped graphene as a function of the applied frequency $\omega$ is demonstrated.\\
\indent
Given this background, the paper is organized as follows. In Sec. 2 we write down the generalized master equations for the average “dephasing” and “depopulation” operators in terms of explicitly time dependent spin-spin relaxation and spin-lattice rates. The latter quantities are expressed in terms of the underlying spectral functions that characterize the electronic and the phonon baths. Numerical plots of these rates are given in this section which demonstrate the transition from the non-Markovian to the
Markovian regime and their T dependencies, which in turn determine the T dependent Mishchenko parameter. This section also presents experimentally accessible inter-band conductivity of the pristine graphene in the Markovian domain. In Sec. 3 we turn to the case of gapped graphene which brings-in the third component of the spin transverse to the graphene layer. The role played by the transverse coupling parameter $\Delta$ in the temperature dependence of the conductivity and an unexpected switching behavior of the latter as a function of $\omega$ are presented here. Section 4 concludes the paper with a summary of our main results.
\section{Model and Method}
In this section we introduce the relevant spin-Boson Hamiltonian for the electric field driven graphene in contact with dissipative Bosonic bath that is modelled as a collection of harmonic oscillators. Applying a unitary transformation in the interaction picture of our system-plus-bath Hamiltonian, we can rephrase the Hamiltonian in the so-called “rotating wave approximation” (RWA) [20]. Since all the rapidly oscillating terms eventually die down in the steady state, we ignore these terms utilizing RWA. Although RWA is a well known tool in quantum optics, its application in the present context of dissipative dynamics of graphene yields a modified spin-boson Hamiltonian which  shapes the foundation of our further study of dissipative dynamics in terms of a master equation for the “reduced” density operator. We also introduce the current density which we will require for computing the nonlinear conductivity beyond the Drude/Kubo regime [23].
\subsection{Model Hamiltonian and Method}
Here we consider graphene as a two band electronic system which is interacting with the surrounding environment. In the Dirac limit, the Hamiltonian of this open system can be written in the system-plus-bath approach of Caldeira-Leggett[24] :
\begin{equation}
H = H_S + H_{SB} + H_{B},
\end{equation}
where, $H_s$ is the subsystem Hamiltonian of the graphene for a given $\bf{k}$ ($\equiv$ to momentum $\bf{p}$),
\begin{equation}
H_S= v_F(\sigma\cdot \textbf{k}),
\end{equation}
where $v_F$ is the fermi velocity. Further, $H_{SB}$ takes into account two distinct physical interactions between the Dirac electron of the graphene and the surrounding phonons and other electrons which can be described by two types of interaction terms: a dissipationless decoherence term related with electron-electron interaction and another dissipative decoherence term caused by electron-phonon interaction. Thus,
\begin{equation}
H_{SB}=\Pi_\textbf{k} X_e + Y_\textbf{k} X_p,
\end{equation}
with the depopulation operator $\Pi_\textbf{k}=(|c_\textbf{k}\rangle\langle c_\textbf{k}|-|v_\textbf{k}\rangle\langle v_\textbf{k}|)$, dephasing operator $Y_k=-i(|v_\textbf{k}\rangle\langle c_\textbf{k}|-|c_\textbf{k}\rangle \langle v_\textbf{k}|)$, $X_e=\sum_q G_q(b_q+b_q^{\dagger})$, and $X_p=\sum_q g_q(a_q+a_q^{\dagger})$. $G_q$ and $g_q$ parameterize the coupling of our system with the surrounding electrons and phonons respectively. $b_q$ and $b_q^{\dagger}$ are the annihilation and creation operators for electrons, while $a_q$ and $a_q^{\dagger}$ are used to denote annihilation and creation operators for phonons. Further, $|c_\textbf{k}\rangle$ and $|v_\textbf{k}\rangle$ are the conduction and valence band eigenfunctions of $H_S$. Finally, the surrounding environment is modelled through usual bosonic structure :
\begin{equation}
H_B = \sum_{q} \omega_q b_q^{\dagger}b_q+\sum_{q}\Omega_q a^{\dagger}_q a_q,
\end{equation}
where the first term indicates the free electron interaction in terms of electron creation ($b_q^{\dagger}$) and
annihilation operators ($b_q$). As shown in Ref.[25,26], the fermionic bath can be `bosonized' as long as we are only interested in the electron-hole excitations near the fermi surface. The last term in Eq.(4) is the collection of harmonic oscillators to represent phononic bath.\\
Now we perturb our entire system with an external alternating electric field $E_0\cos({\omega t})$. So the new system hamiltonian is
\begin{equation}
H_0 =H_S+H_\omega(t)=H_S+v_F\frac{eE_0}{\omega}\sigma_x\sin{\omega t},
\end{equation}
where $\sigma_x$ is the Pauli spin matrix.
\subsection{Method of Quantum Dynamics}
To study the dynamics of our system we follow the method introduced in Ref.[20]. Our starting point is the Schr$\ddot{o}$dinger picture von Neumann–Liouville equation for the density operator $\rho(t)$ :
\begin{equation}
i\frac{d\rho(t)}{dt}= [H_S+H_{\omega}+H_{SB}+H_B,\rho(t)].
\end{equation}
Now, going into the interaction picture, tracing out the bath degrees of freedom, and utilizing a cumulant expansion scheme one can obtain convolution-less master equation for the reduced density operator $\rho'_S(t)=exp(iH_St)\rho_S exp(-iH_St)$ :
\begin{equation}
    \dfrac{d}{dt}\rho'_S(t)=-i[H^{eff}_\omega(t),\rho'_S(t)]-R(t)\rho'_S(t),
\end{equation}
where $R(t)$ is known as the 'relaxation matrix'.
\begin{equation}
        \begin{split}
            R(t)\rho'_S(t)&=\int_{0}^{t} Tr_B[H_{SB}^I(\tau)H_{SB}^I(0)\rho_B\rho'_S(t)+\\
            &\rho_B\rho'_S(t)H_{SB}^I(0)H_{SB}^I(\tau)+H_{SB}^I(\tau)\rho_B\rho'_S(t)H_{SB}^I(0)\\
            &-H_{SB}^I(0)\rho_B\rho'_S(t)H_{SB}^I(\tau)] \ d\tau.
        \end{split}
\end{equation}
Under rotating wave approximation, our `effective' ac term can be expressed as
\begin{equation}
        H^{eff}_\omega (t) =\Omega_\textbf{k}[Y^{+}_\textbf{k}\exp[-i(\omega-\Delta_\textbf{k})t]+ H.c.]/2,
\end{equation}
where
\begin{eqnarray}
        Y^{+}_\textbf{k}&=&|c_\textbf{k}\rangle \langle v_\textbf{k}|,\nonumber\\
        Y^{-}_\textbf{k}&=&|v_\textbf{k}\rangle \langle c_\textbf{k}|,\nonumber\\
       \Omega_\textbf{k}&=&(eEv_F/\omega) \sin{\chi_\textbf{k}},\nonumber \\
    \Delta_\textbf{k}&=&2v_F|\textbf{k}|.
\end{eqnarray}
Here $\Delta_\textbf{k}$ is the ``tunneling frequency" between the valance and conduction bands and $(\omega-\Delta_\textbf{k})$ is called the ``detuning frequency". On the other hand, the time evolution of $H_{SB}$ in the interaction picture is given by
\begin{eqnarray}
    \begin{split}
        H_{SB}^I(t)&=U^+_S(t) H_{SB} U_S(t)\\
        &=\Pi_\textbf{k}X_e(t)+[-i\exp({i\Delta_\textbf{k} t})Y^{+}_\textbf{k}+\\
        &i\exp({-i\Delta_\textbf{k} t})Y^{-}_\textbf{k}]X_p(t),
    \end{split}
\end{eqnarray}
where $U_S(t)=\exp{-i(H_{S}+H_{B})t}$.
Since our focus is to analyze optical conductivity of graphene, we can introduce an average momentum-resolved current density along the applied electric field  as follows (explained in detail in sec.IID) :
\begin{eqnarray}
j_{\bf{k}x}(t)=ev_F[\cos(\chi_{\bf{k}})\langle \Pi_{\bf{k}}(t)\rangle+\sin(\chi_{\bf{k}})\langle Y_{\bf{k}}(t)\rangle],
\end{eqnarray}
where $e$ is the electronic charge, and $\chi_{\bf{k}}$ is the angle between the $\bf{k}$ vector and the $x$ axis. Hence our task is to calculate $\langle \Pi_{\bf{k}}\rangle$ and $\langle Y_{\bf{k}}\rangle$ (where $\langle...\rangle$ represents expectation values) to obtain nonlinear optical conductivity from the average current density expression. Now, invoking Markov approximation, one can extend the upper limit of the integral to infinity rendering the relaxation matrix $R(t)$. After some algebra, one can write the equation which governs the dynamics of $\langle \Pi_{\bf{k}}(t)\rangle$:
\begin{eqnarray}
\begin{split}
\frac{d\langle \Pi_{\bf{k}}(t)\rangle}{dt} &= i\Omega_\textbf{k}[\langle Y^{-}_\textbf{k}\rangle\exp(i\omega t)-\langle Y^{+}_\textbf{k}\rangle\exp(-i\omega t)]\\
&-\gamma_p[\langle\Pi_\textbf{k} (t)\rangle-\langle\Pi_\textbf{k}\rangle_{eq}]],
\end{split}
\end{eqnarray}
and the equation which governs the time evolution of $\langle Y^+_{\bf{k}}\rangle$ is
\begin{eqnarray}
\begin{split}
\frac{d\langle Y^+_{\bf{k}}(t)\rangle}{dt}&=-\dfrac{i\Omega_\textbf{k}}{2}\langle\Pi_\textbf{k}\rangle \exp(i\omega t)-\\
&(\gamma_e-i\Delta_\textbf{k})\langle Y^{+}_\textbf{k}\rangle.
\end{split}
\end{eqnarray}
One can obtain the value of $Y^-_{\bf{k}}$ by taking the complex conjugate of $Y^+_{\bf{k}}$ . Here the spin-lattice relaxation rate $\gamma_p$ and the spin-spin relaxation rate $\gamma_e$ are given as follows :
\begin{equation}
\gamma_p = 2\int_{-\infty}^{\infty}d\tau \cos(\Delta_\textbf{k}\tau)\xi_p(\tau),
\end{equation}
with the phonon bath correlation function
\begin{equation}
\xi_p(t) =\int_{0}^{\infty}d\omega J_p(\omega)[\coth\Big(\frac{\beta\omega}{2}\Big)\cos(\omega t)-i\sin(\omega t)],
\end{equation}
$J_p(\omega)$ being the so-called phonon spectral function. On the other hand, the spin-spin relaxation rate is given by :
\begin{equation}
\gamma_e=2\int_{-\infty}^{\infty}d\tau\int_{0}^{\infty}d\omega  J_e(\omega)[\coth\Big(\frac{\beta\omega}{2}\Big)\cos(\omega t)-i\sin(\omega t)],
\end{equation}
where the spectral function for the electronic bath is given by $J_e(\omega)$.
\subsection{Detailed inspection of $\gamma_e$ and $\gamma_p$}
In an earlier phenomenological treatment [19], the relaxation rates are considered coarse-grained, frequency independent and temperature independent phenomenological constants. Further, they analyse the steady state electrical response in different linear and nonlinear regime within the Markovian approximation of the phenomenological rate equation. However, as mentioned earlier, the present spin-boson model is a microscopic theory that adopt the machinery of nonequilibrium statistical mechanics. As a result, the genesis of the relaxation rates can be connected to the details of the spectral fluctuations of the underlying phonon and electron baths. It is the goal of this subsection to provide an extensive analysis of the interaction of the system with surrounding thermal and electronic baths. The variation of these rates with temperature is also investigated.\\
\begin{figure}[h]
	\centering
	\includegraphics[width=4.2cm,height=3.2cm]{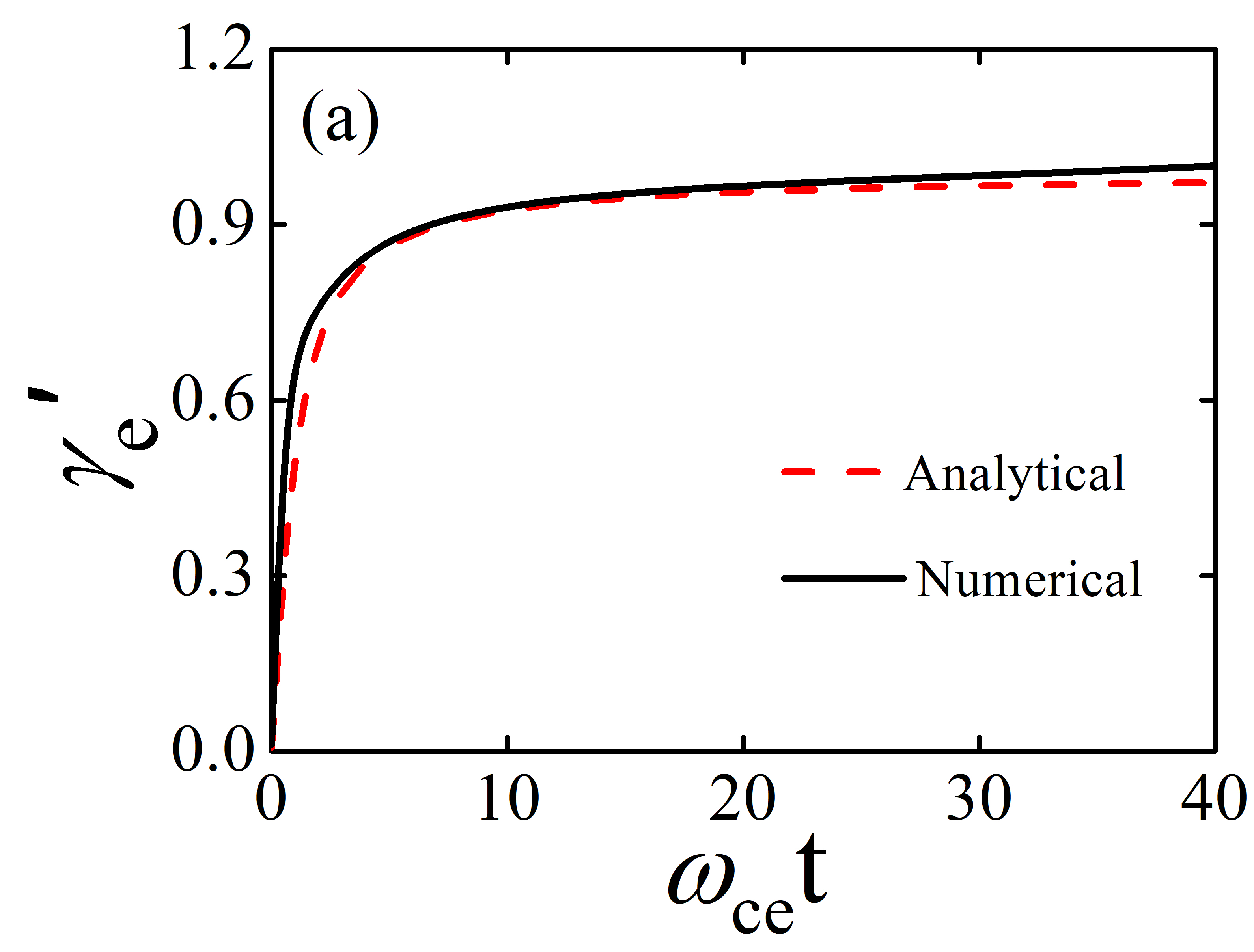}
        \includegraphics[width=4.2cm,height=3.0cm]{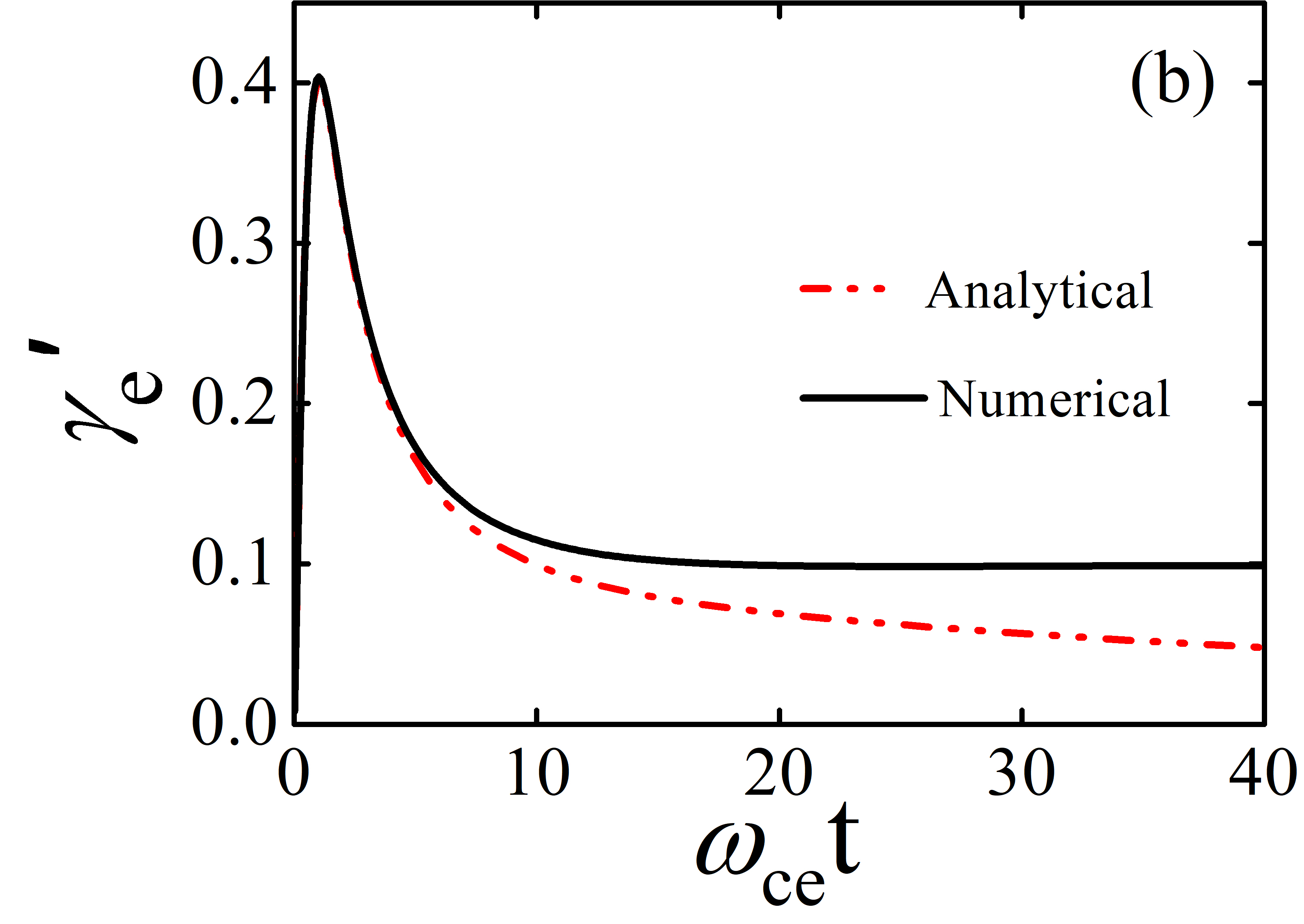}
        \includegraphics[width=4.2cm,height=3.2cm]{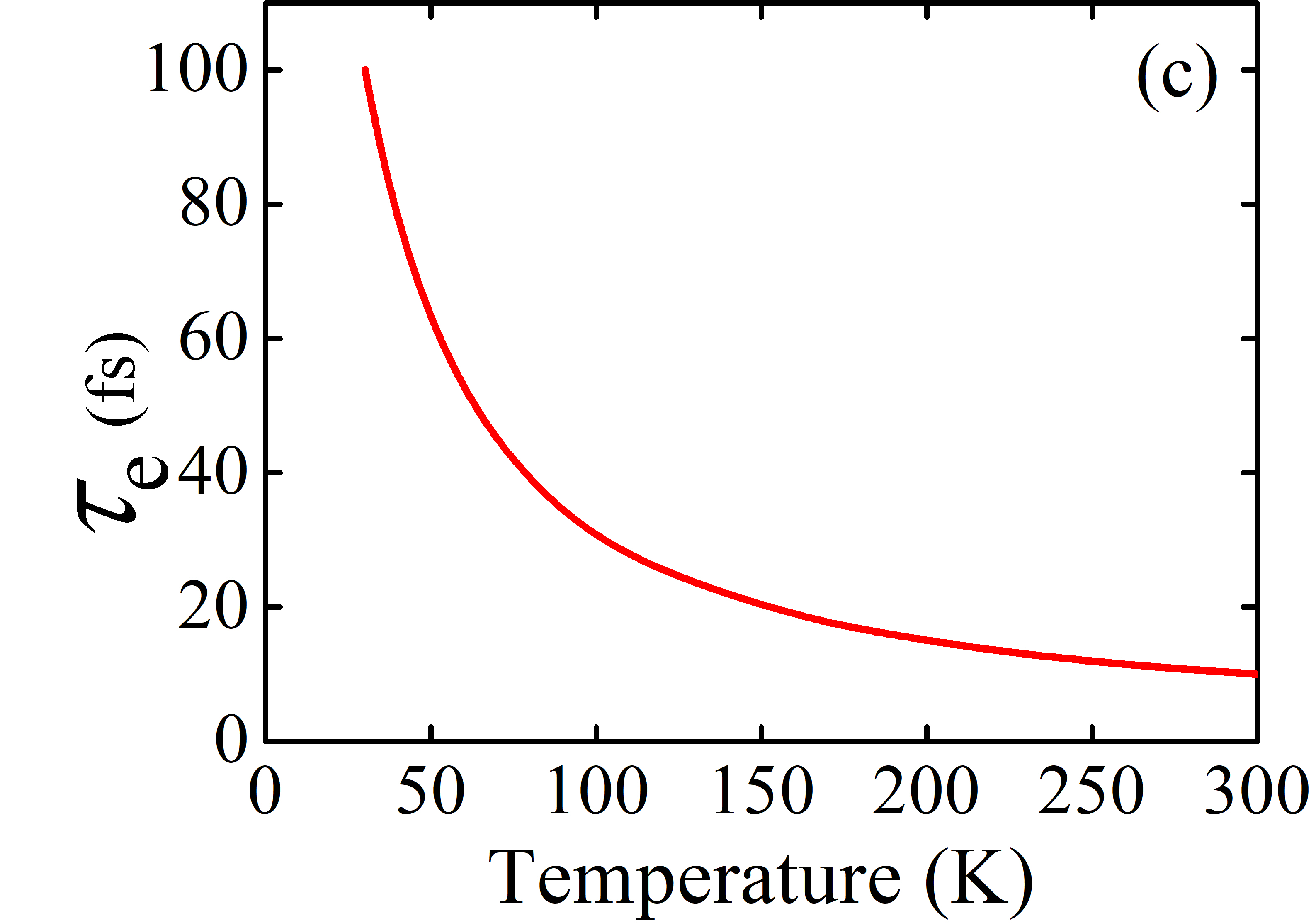}
	\caption {Variation of dimensionless spin-spin relaxation rate $\gamma^{'}_e$ ($\gamma^{'}_e={\gamma_e}/{\omega_{ce}}$) with dimensionless time $\omega_{ce} t$ at (a) 300 K  and (b) 30 K temperature, respectively. In both the cases analytical results (red dashed line) are compared with the numerically simulated results (solid black line). The variation of spin-spin relaxation time ($\tau_e$) with temperature is shown in (c). For the plotting purpose we use $\omega_{ce} = 10^{14}$ Hz and $\alpha_{e} = 0.2$.}
	\label{EEDl}
\end{figure}
The time dependency of these relaxation rates demonstrating the transition from non-Markovian domain to Markovian domain can be expressed by modifying the Eq.(17) and Eq.(15).Thus,
\begin{equation}
    \gamma_{e}(t)=4\int^{t}_0 d\tau\int^{\infty}_0J_e(\omega){\coth\Big(\frac{\beta\omega}{2}\Big)}\cos{\omega\tau} d\omega,
\end{equation}
where $J_e(\omega)$ is the Ohomic spectral function for electron bath with exponential cutoff frequency $\omega_{ce}$:
\begin{equation}
    J_e(\omega)=\alpha_{e}\omega\exp(-{\omega}/{\omega_{ce}}),
\end{equation}
where $\alpha_{e}$ is a coupling parameter.
Thus,
\begin{equation}
    \gamma_{p}(t)=4\int^{t}_0 d\tau \cos{(\Delta_{\textbf{k}}\tau)}\int^{\infty}_0J_p(\omega)\coth{\Big(\frac{\beta\omega}{2}\Big)}\cos{(\omega\tau)} d\omega,
\end{equation}
where $J_p(\omega)$ has the usual Debye structure[21] with cutoff frequency $\omega_{cp}$,
\begin{equation}
    J_p(\omega)=\alpha_{p}\dfrac{\omega^3}{\omega^2_{cp}}\exp{(-{\omega}/{\omega_{cp}})},
\end{equation}
with $\alpha_{p}$ is the coupling parameter.\\
\indent
In the continuation of the above discussion, we can now derive the closed form expressions of $\gamma_e(t)$ and $\gamma_P(t)$ for the high-T as well as in the low-T regime. In the high-T regime ($\beta\omega_{ce}<< 1, \coth(\frac{\beta\omega}{2})\approx \frac{2}{\beta\omega}$) one can obtain the relaxation rate $\gamma_e(t)$ in the following form:
\begin{eqnarray}
\gamma_e(t)\simeq 8\alpha_e k_BT \tan^{-1}(-\omega_{ce}t).
\end{eqnarray}
Figure (1a) shows the comparison between the numerically simulated results (black solid line) and the analytical results (Eq. (22); red dashed line) of the variation of dimensionless spin-spin relaxation rate ($\gamma^{'}_e$ ) as a function of dimensionless time ($\omega_{ce}t$) for 300 K. Both results fairly match with each other.
On the other hand, low-T behaviour of $\gamma_e(t)$ can be obtained by using the relation $\coth(\frac{\beta\omega}{2})\approx [1+2\exp(-\beta\omega)]$. The low-T expression of $\gamma_e(t)$ is given by
\begin{equation}
\gamma_e(t)=4\alpha_{e}\omega_{ce}[I_1+2I_2],
\end{equation}
where $I_1= \dfrac{\omega_{ce} t}{(\omega_{ce} t)^2 +1}$ and $I_2=\dfrac{\omega_{ce} t}{(\omega_{ce} t)^2 +(1+(\beta \omega_{ce}))^2}$. We have observed close agreement of these analytical results (red dashed line) with that of numerically simulated outcomes (solid black line) in Figure (1b).\\
\indent
A distinct transient region of spin-spin relaxation rate is observed for both plots corresponding to the two temperature values. The transient region is generally known as the ‘Non-Markovian regime, which occurs at a shorter time scale than the quantal time ($\hbar/{k_{B}T}$). In Markovian approximation technique, all the quantum phenomena occurring within this particular quantal time scale can be neglected. The values of the quantal time scale are much shorter than the spin-spin relaxation time at the non-transient region for both temperatures. It is evident from the figure that the spin-spin relaxation time ($\tau_e=\frac{1}{\gamma_e}$) for pristine graphene is 10 fs (in Markovian region) at 300 K, which directly supports the experimentally obtained values of spin-spin relaxation time reported earlier [27]. The $\tau_e$ is increased to $100$ fs at $30$ K, inferring the strong temperature-dependent nature of the $\tau_e$. The variation of  $\tau_e$ with temperature is demonstrated in Figure (1c).\\
\indent
\begin{figure}[h]
	\centering
	\includegraphics[width=4.5cm,height=3.5cm]{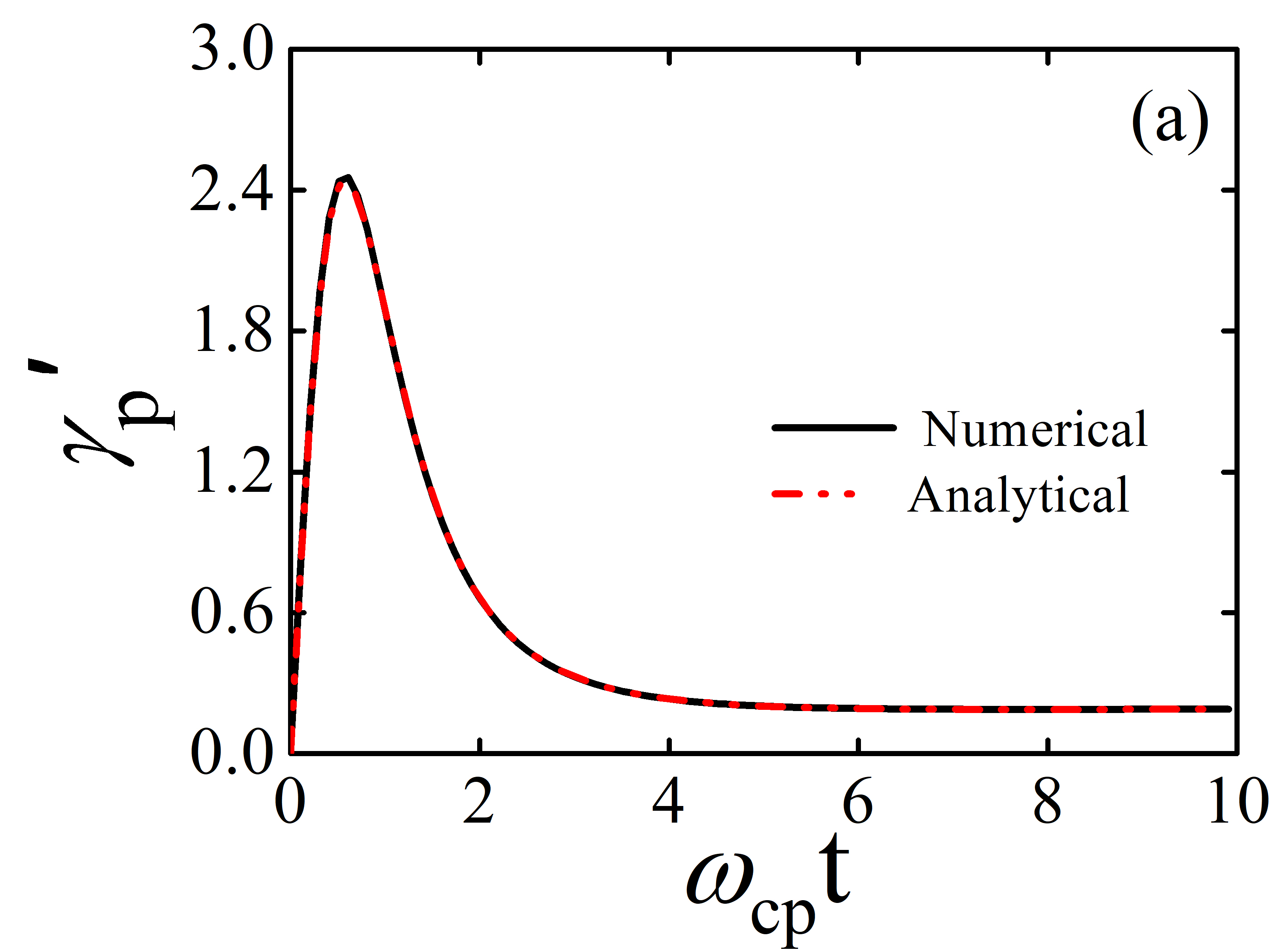}
     \includegraphics[width=4.2cm,height=3.1cm]{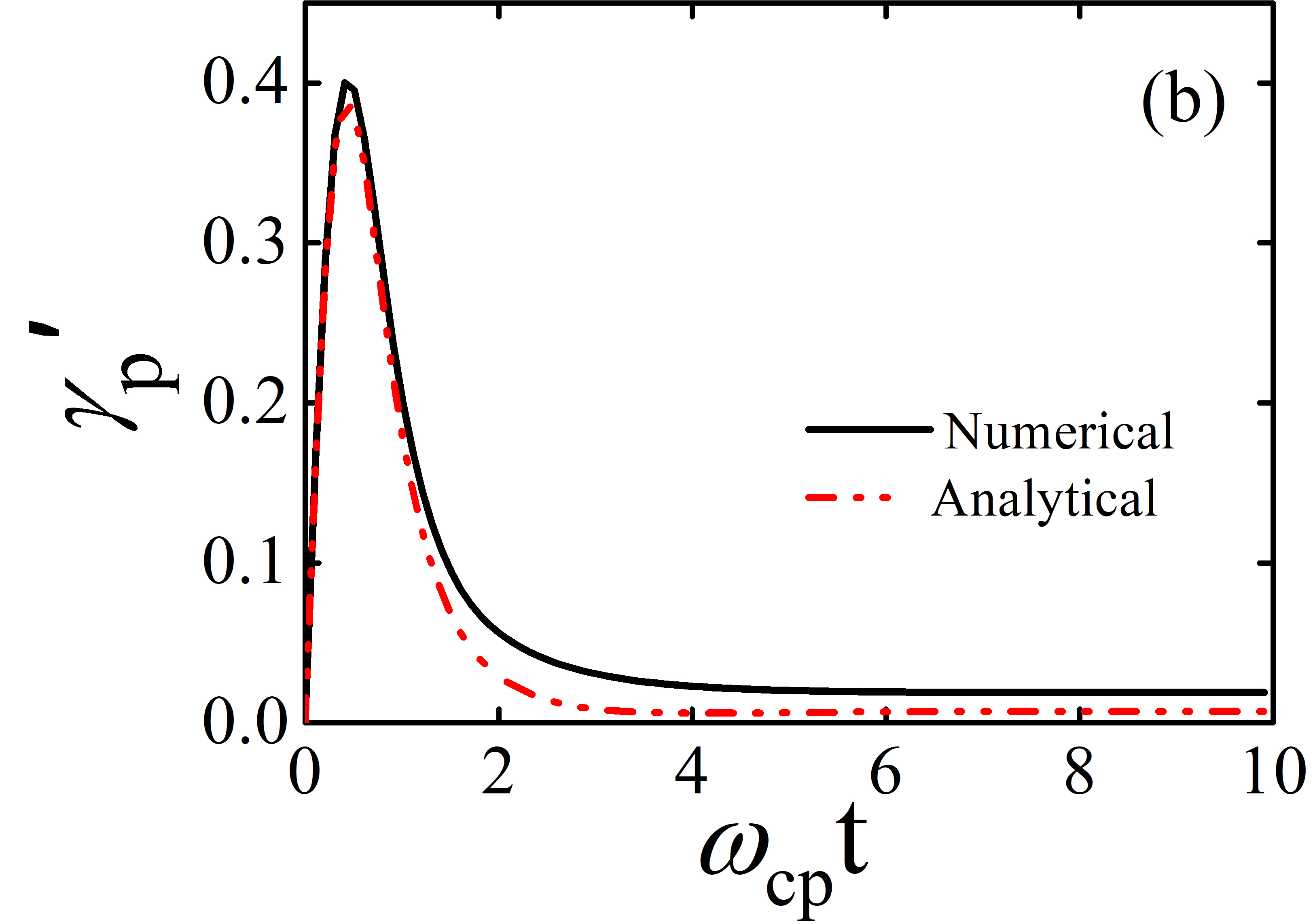}
        \includegraphics[width=4.2cm,height=3.2cm]{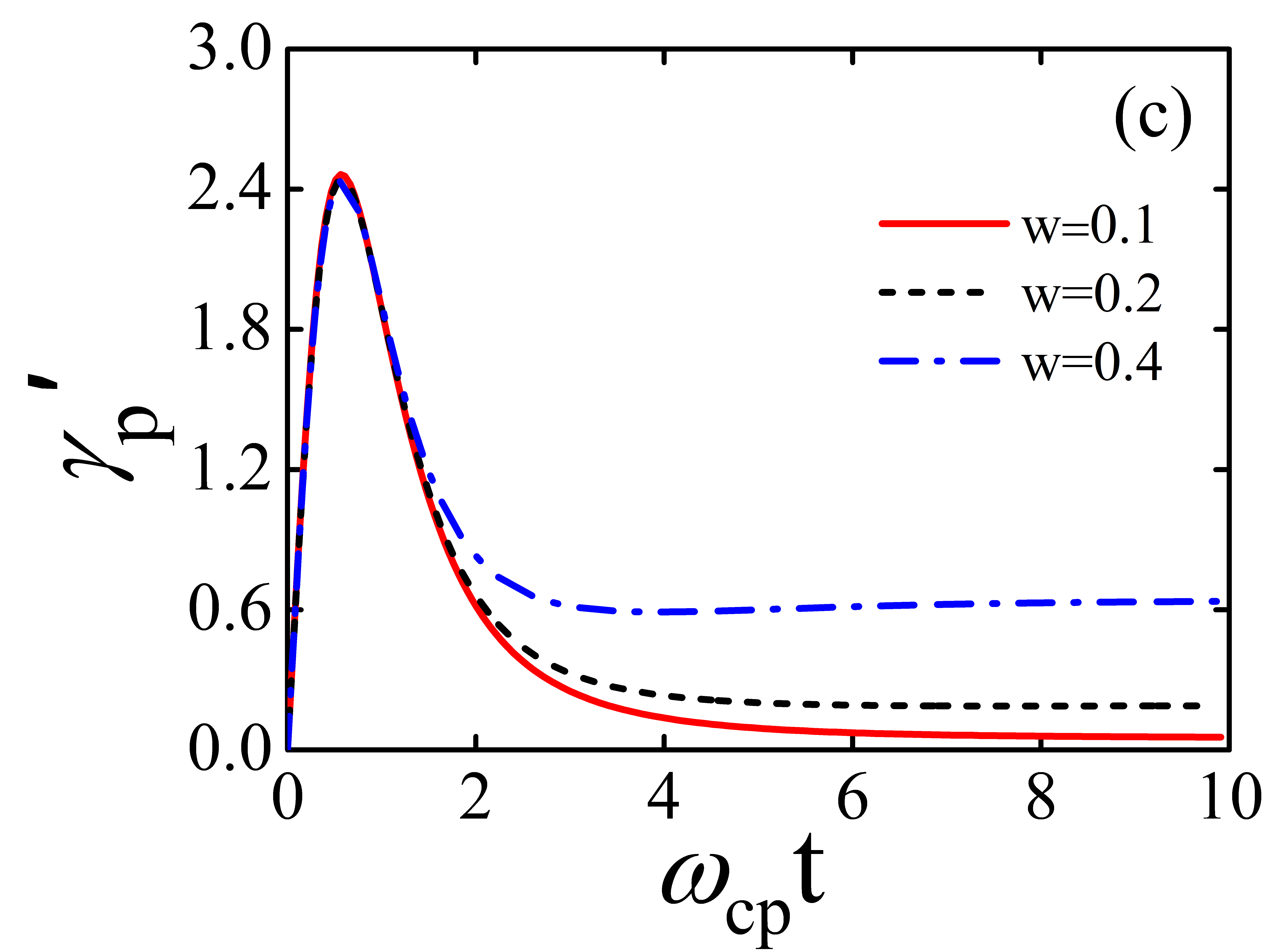}
        \includegraphics[width=4.0cm,height=3.2cm]{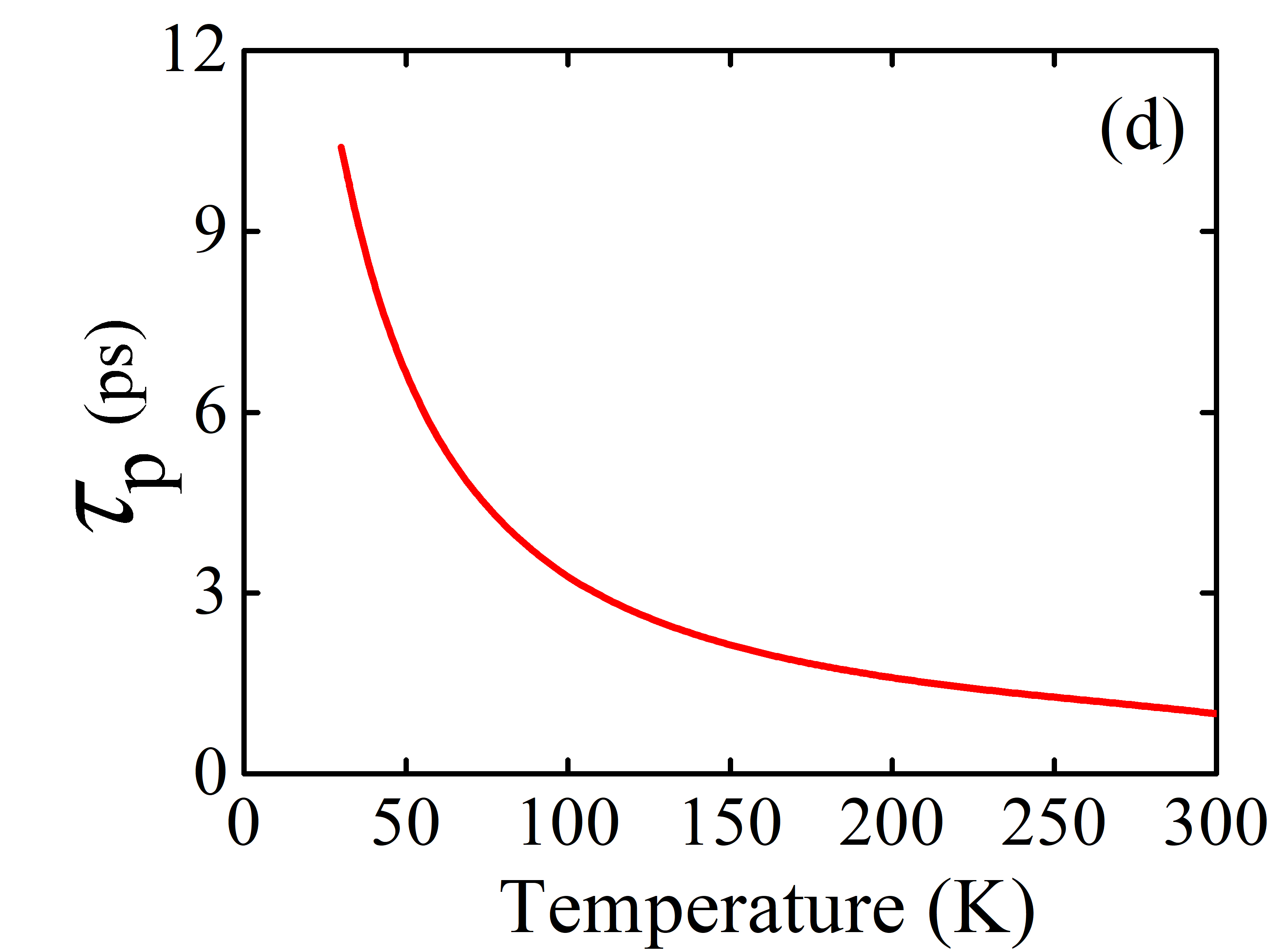}
        \includegraphics[width=4.5cm,height=3.5cm]{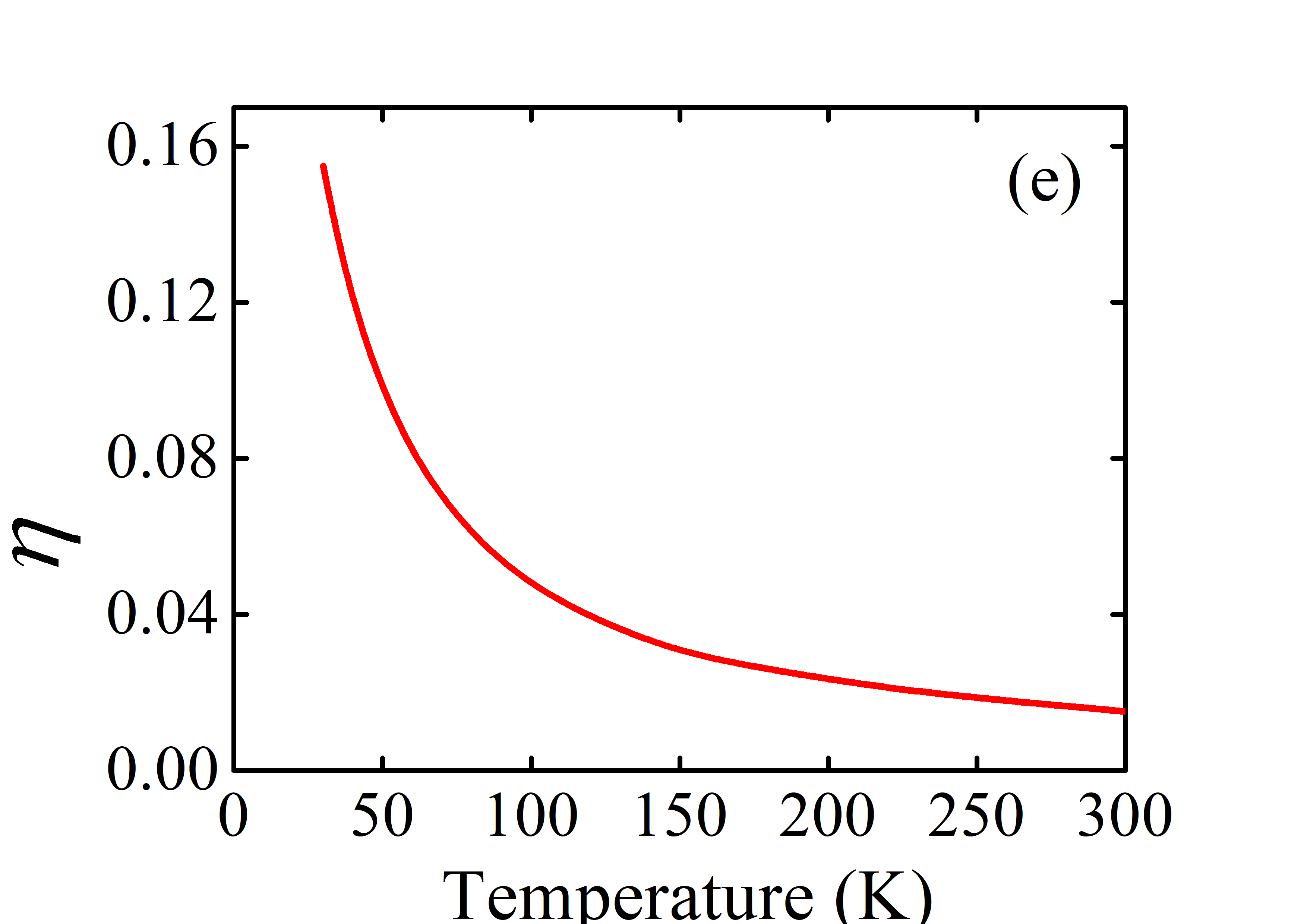}
	\caption {Variation of dimensionless spin-lattice relaxation rate $\gamma^{'}_p$ ($\gamma^{'}_p={\gamma_p}/{\omega_{cp}}$) with dimensionless time $\omega_{cp} t$ at (a) 300 K and (b) at 30 K temperature, respectively. In both cases analytical results (red dashed line) are compared with the numerically simulated results (solid black line). For the plotting purpose we used $\omega_{cp} = 5\times10^{12}$ Hz and $\alpha_{p}$ = 0.06. (c) Profile of dimensionless spin-lattice relaxation rate $\gamma^{'}_p$ for different values of detuning frequency $w$ ($w={\Delta_{\textbf{k}}}/{\omega_{cp}})$. (d) Plot of spin-lattice relaxation time $\tau_p$ as a function of temperature. (e)Plot of Mischenko parameter ($\eta$) as function of temperature. }
	\label{EEDl}
\end{figure}
Let us move to the analysis of the spin-lattice relaxation that is considered as the key process for interband transition in our work, which can be described by Eq.(20).At high temperature, utilizing $\coth(\frac{\beta\omega}{2})\approx \frac{2}{\beta\omega}$ and considering Eqs.(20) and (21), one can show
\begin{equation}
\gamma_p(t)={16\alpha_pk_BT}\int_{0}^{\omega_{cp}t}dx \cos(bx)\frac{[1-3x^2]}{[1+x^2]^3},
\end{equation}
where $b=\frac{\Delta_{\mathbf{k}}}{\omega_{cp}}$ and $x=\omega_{cp}\tau$. The closed form expression of $\gamma_p(t)$ is given in Appendix A (See Eq. (A.2)). The variation of dimensionless spin-lattice relaxation rate ($\gamma^{'}_p$) with dimensionless time ($\omega_{cp} t$) for high-T i.e. at 300 K is plotted in Figure (2a).
The close agreement between this analytical expression (Eq. (24), red dashed line) with the numerically simulated results (black solid line) is demonstrated in Figure (2a). On the other hand, at low temperature :
\begin{eqnarray}
&&\gamma_p(t)=24\alpha_p\omega_{cp}\int_{0}^{\omega_{cp}t}dx\frac{[1-6x^2+x^4]}{[1+x^2]^4}\cos(bx) \nonumber \\
&+&2 \int_{0}^{\omega_{cp}t}dx \frac{[(1+a_p)^4-6(1+a_p)^2x^2+x^4]}{[(1+a_p)^2+x^2]^4}\cos(bx), \nonumber \\
\end{eqnarray}
with $a_p=\frac{\omega_{cp}}{k_BT}$. We compare our analytical expression (Eq. (25)) with that of numerical results in Figure (2b). Figures (2a) and (2b) both exhibit a transition of the spin-lattice relaxation rate from  non-Markovian region to Markovian regime, similar to the electron induced relaxation rate as mentioned earlier. In the Markonivan region, the phonon induced relaxation time ($\tau_p=\frac{1}{\gamma_p}$) of graphene is found to be 1 ps at 300 K, which is further increased to 10 ps at 30 K. The calculated values of spin-lattice relaxation time fairly agree with previously reported experimental values [28]. It is evident from Eq.(25) that the value of spin-lattice relaxation time significantly depends on $\Delta_{\textbf{k}}$ (tunneling frequency), which is related to detuning frequency. Hence we also show the variation of dimensionless spin-lattice relaxation rate for different $\Delta_{\textbf{k}}$ values at 300 K temperature in Figure (2c). In the Markovian region, the spin-lattice relaxation rate is increased with increasing $\Delta_{\textbf{k}}$ value. The spin-lattice relaxation time are 3.6 ps,1 ps, and 320 fs for w=0.1, 0.2 and 0.4, respectively (Figure 2c). Higher value of $\Delta_{\textbf{k}}$ is associated with the carriers having high momentum ‘k’ value, which relax (interband) faster by interacting with the lattice and results faster spin-lattice relaxation time of the carriers. The temperature dependent spin-lattice relaxation time is plotted in Figure (1d) which confirms the increasing nature of the spin-lattice relaxation time at low temperatures compared to its high temperature values. The deficiency of phonon scatterers at low temperatures may enhance the spin-lattice relaxation time of the carriers at the conduction band. Inspired by the previous studies[19], the Mischenko parameter ($\eta$) can be utilized to explain the nonlinear optical response of graphene. Hence, our study predicts strong temperature dependency of the both electron induced and phonon induced relaxation processes, which makes the Mischenko parameter a function of temperature. In Figure (2e), we demonstrate the temperature dependency of the Mischenko parameter.\\
\subsection{Nonlinear optical conductivity : Pristine Graphene}
 The linear and nonlinear response of the graphene system can be quantified by this single dimensionless parameter $\eta$, where $\eta<<1$ describes the linear regime and $\eta>>1$ denotes the nonlinear regime. Typically, one can divide the optical conductivity in four distinct regimes: (a) linear response in the clean regime, (b) linear response in the dirty regime, (c) nonlinear response in the clean regime, and finally, (d) nonlinear response in the dirty regime. Here, we are denoting the clean (dirty) regime as the collisionless or high-frequency limit (collisional or low frequency), and this can be quantified by the region $\frac{\gamma^{st}_{e}}{\omega} << 1$ ($ \frac{\gamma^{st}_e}{\omega }\geq 1$) as the steady state value of $\gamma_e(t)$. We consider $\gamma_e=\gamma_e^{st}$ for further discussion. \\
\indent
In order to study the conductivity of a system, we need to calculate the steady state current density operator in the direction of the applied electric field. For simplicity, we consider that the frequency dependent electric field is applied along the $x$ axis and the response to the field is measured after the system attains the steady state. In general, the nonlinear response has a component in-phase with the applied field and another out-of-phase with it. To proceed further let us introduce the current density operator :
\begin{equation}
{\bf{j}}(t)=-\frac{g_sg_{v}}{(2\pi)^2}\int d{\bf{k}} j_{\bf{k}}(t),
\end{equation}
where $g_s$ ($g_{v}$) is the spin (valley) degeneracy factor (in our case, both are 2), the momentum dependent component of particle current density is $j_{\bf{k}}(t)=eTr[\rho_{\textbf{k}}(t)\vec{v}_k(t)]$. Since the electric field is applied in the $x$ direction, the $x$ direction component of the momentum dependent current density in the steady state is given by :
\begin{equation}
j_{\textbf{k}x}(t)_{st}= ev_F[\cos(\chi_\textbf{k})\langle\Pi_\textbf{k}(t)\rangle_{st}+\sin(\chi_\textbf{k})\langle Y_\textbf{k}(t)\rangle_{st}],
\end{equation}
where the first term carries the contributions from the intraband transitions, while the second term includes the effect of interband contributions. It is observed that if one summed over all $\vec{\textbf{k}}$ vectors the intraband term does not contribute to the optical conductivity in graphene [20]. Following Ref.[20], one can obtain the steady state expressions of $\langle \Pi_\textbf{k}(t)\rangle$ and $\langle Y_{\bf{k}}^{+}(t)\rangle$ from Eqs. (13) and (14). Thus we can obtain the steady state form as:
\begin{eqnarray}
\langle\Pi_\textbf{k}\rangle_{st}=\langle\Pi_\textbf{k}\rangle_{eq}[1+\dfrac{\dfrac{\gamma_e}{\gamma_p}\Omega^{2}_\textbf{k}}{\gamma^{2}_e+(\Delta_\textbf{k}-\omega)^{2}}]^{-1},
\end{eqnarray}
and
\begin{eqnarray}
      \langle Y_\textbf{k}\rangle_{st}&=-\dfrac{\Omega_\textbf{k}\langle\Pi_\textbf{k}\rangle_{st}[\gamma_e\cos{\omega t}+(\Delta_\textbf{k}-\omega)\sin{\omega t}]}{(\Delta_\textbf{k}-\omega)^{2}+\gamma^{2}_e}.
\end{eqnarray}
It is well known that only the in-phase term of $\langle Y_k(t)\rangle_{st}$ contributes to the dissipative component of optical conductivity. Thus, the general expression for the nonlinear optical conductivity is given by :
\begin{eqnarray}
    \sigma_{xx}=\dfrac{g_sg_\nu}{(2\pi)^{d}}\int \dfrac{e^{2}[v_F\sin{\chi_\textbf{k}}]^{2}\langle\Pi_\textbf{k}\rangle_{st}\gamma_e}{\omega((\Delta_\textbf{k}-\omega)^{2}+\gamma^{2}_e)}d\textbf{k}.
\end{eqnarray}
In light of the above discussion, the optical conductivity can be categorized into four regimes(lc,ld, nc, nd) by rewriting:
\begin{equation}
    \begin{aligned}
      \dfrac{\gamma_e}{\gamma_p}\Omega^{2}_\textbf{k}&=[\eta\gamma_e\sin{\chi_\textbf{k}}]^{2}.\\
    \end{aligned}
\end{equation}
Let us now proceed to further discussion of the longitudinal optical conductivity in detail for all the four regimes.
\subsubsection{Linear clean regime : ($\eta << 1, \frac{\gamma_e}{\omega} << 1$)}
    In this regime we retain the zeroth order of $\eta$ which enables us to consider $\langle\Pi_\textbf{k}\rangle_{st}\approx \langle\Pi_\textbf{k}\rangle_{eq}$ in Eq.(28) and one may convert the Lorentzian part of (30) into a Dirac-delta function. Thus we obtain
    \begin{equation}
       \sigma_{xx}=\dfrac{\pi g_sg_v e^2 v^{2}_F}{\omega (2\pi)^{2}}\int\sin^{2}{\chi_\textbf{k}} \delta(\Delta_\textbf{k}-\omega) (f_{ck}-f_{vk}) d\textbf{k},
    \end{equation}
    where, $f_{ck}(f_{vk})$ is the Fermi-Dirac distribution function for the conduction (valence) band. For graphene, the system has particle-hole symmetry and isotropic quasi-particle dispersions which enables one to  write,
    \begin{eqnarray}
       \sigma_{xx} &=& \dfrac{\pi g_sg_v e^2 v^{2}_F g(\omega,\mu,T)}{\omega (2\pi)^{2}}\int\sin^{2}{\chi_\textbf{k}} \delta(\Delta_\textbf{k}-\omega)\,d\textbf{k}\nonumber\\
       &=&\dfrac{e^2 g(\omega,\mu,T)}{4},
    \end{eqnarray}
    where
\begin{equation}
 g(\omega,\mu,T)=\dfrac{1}{2}[\tanh{\dfrac{\omega+2\mu}{4k_{B}T}}+\tanh{\dfrac{\omega-2\mu}{4k_{B}T}}],
 \end{equation}
 where $\mu$ is the chemical potential of the system. In the limt $T\to 0$, the function $g(\omega,\alpha,T\to 0)=\Theta(\dfrac{\omega}{2}-|\alpha|)$ \cite{ref16},with $\Theta(x)$ as the Heaviside step function . As a result the conductivity at zero temperature becomes
            \begin{equation}
        \begin{aligned}
          \sigma_{xx}
          &=\dfrac{e^2}{4}\Theta(\dfrac{\omega}{2}-|\mu|).
        \end{aligned}
    \end{equation}
\subsubsection{Nonlinear clean regime :($\eta\geq 1, \frac{\gamma_e}{\omega} << 1$)}
 In the nonlinear clean limit, one may typically consider the Lorentzian by a Dirac delta function, and Eq. (30) reduces to
   \begin{equation}
   \begin{aligned}
       \sigma_{xx}&=\dfrac{\pi g_sg_\nu e^2 v^{2}_F}{\omega (2\pi)^{2}}\int \dfrac{\sin^{2}{\chi_\textbf{k}} \delta(\Delta_\textbf{k}-\omega)}{1+\eta^2\sin^{2}{\chi_\textbf{k}}} (f_{ck}-f_{vk})\,d\textbf{k}\\
         &=\dfrac{e^2 g(\omega,\mu,T)}{4} \dfrac{2}{\eta^2} \Big[1-\dfrac{1}{\sqrt{(1+\eta^2)}}\Big].
       \end{aligned}
   \end{equation}
   As $T\to 0$ the conductivity reduces to
 \begin{equation}
   \begin{aligned}
       \sigma_{xx}&=\dfrac{e^2}{4} \dfrac{2}{\eta^2} \Big[1-\dfrac{1}{\sqrt{(1+\eta^2)}}\Big]\Theta(\dfrac{\omega}{2}-|\mu|).
       \end{aligned}
   \end{equation}

\begin{figure}[h]
	\centering
	\includegraphics[width=4.2cm,height=3.5cm]{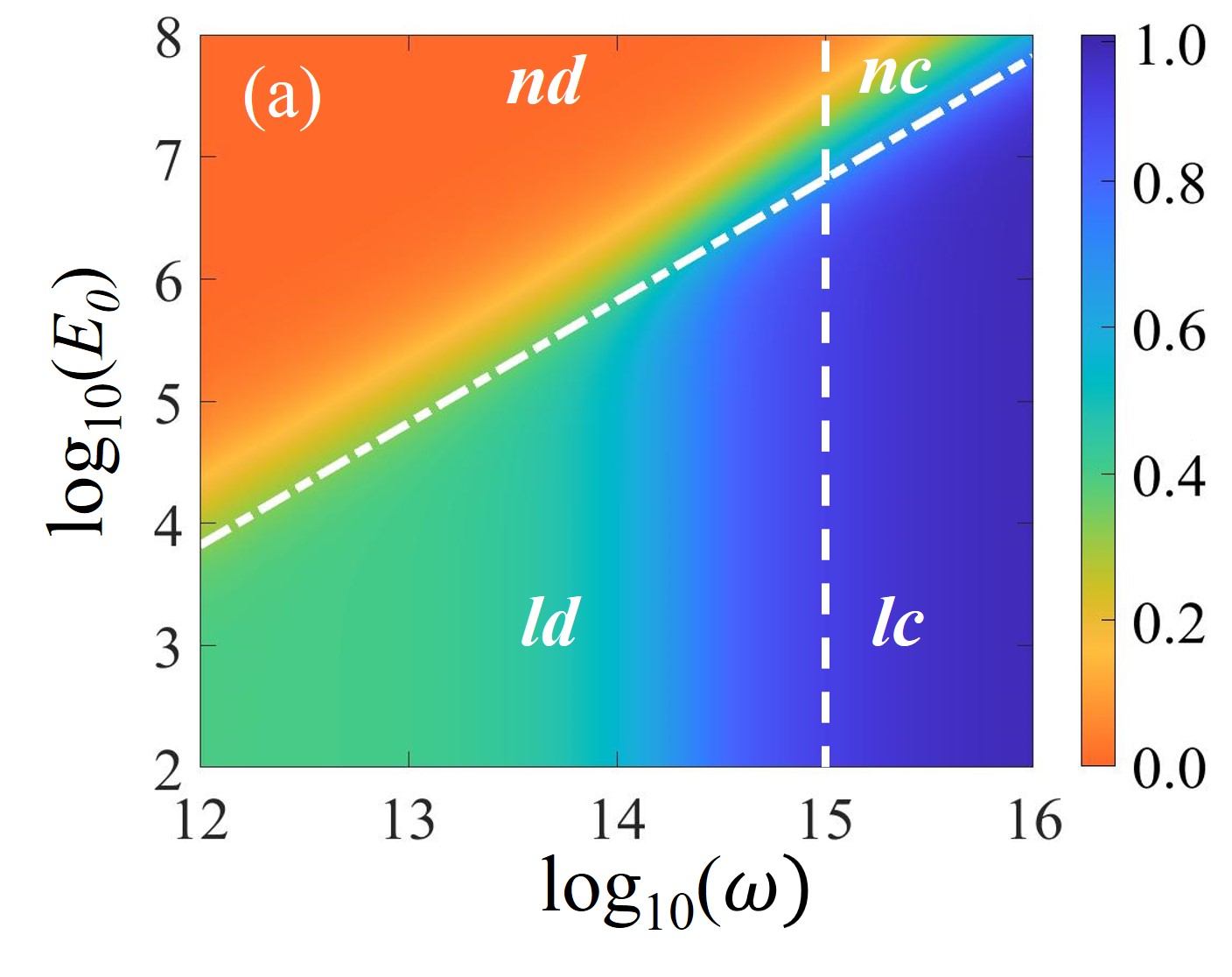}
        \includegraphics[width=4.2cm,height=3.5cm]{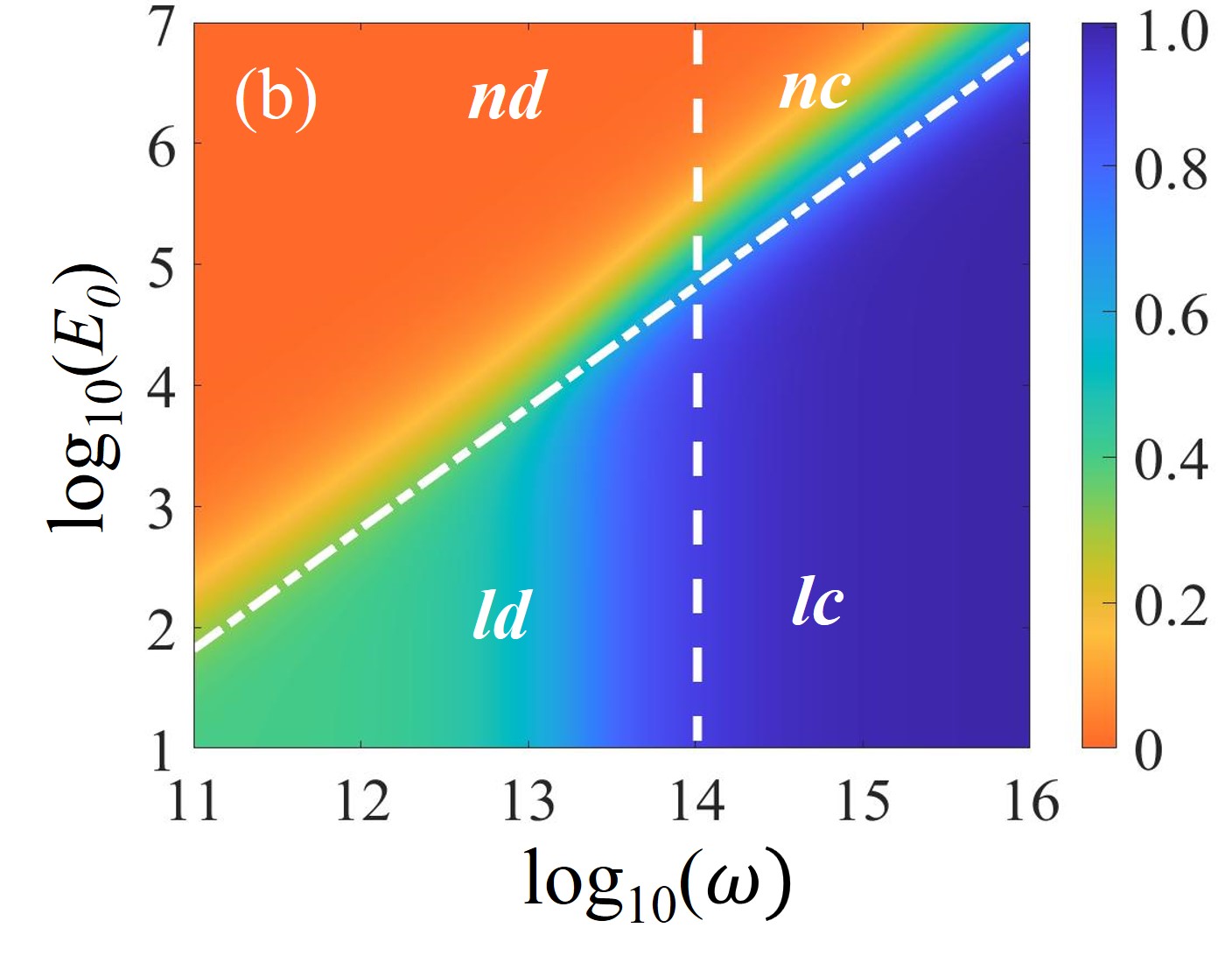}
        \includegraphics[width=4.2cm,height=3.5cm]{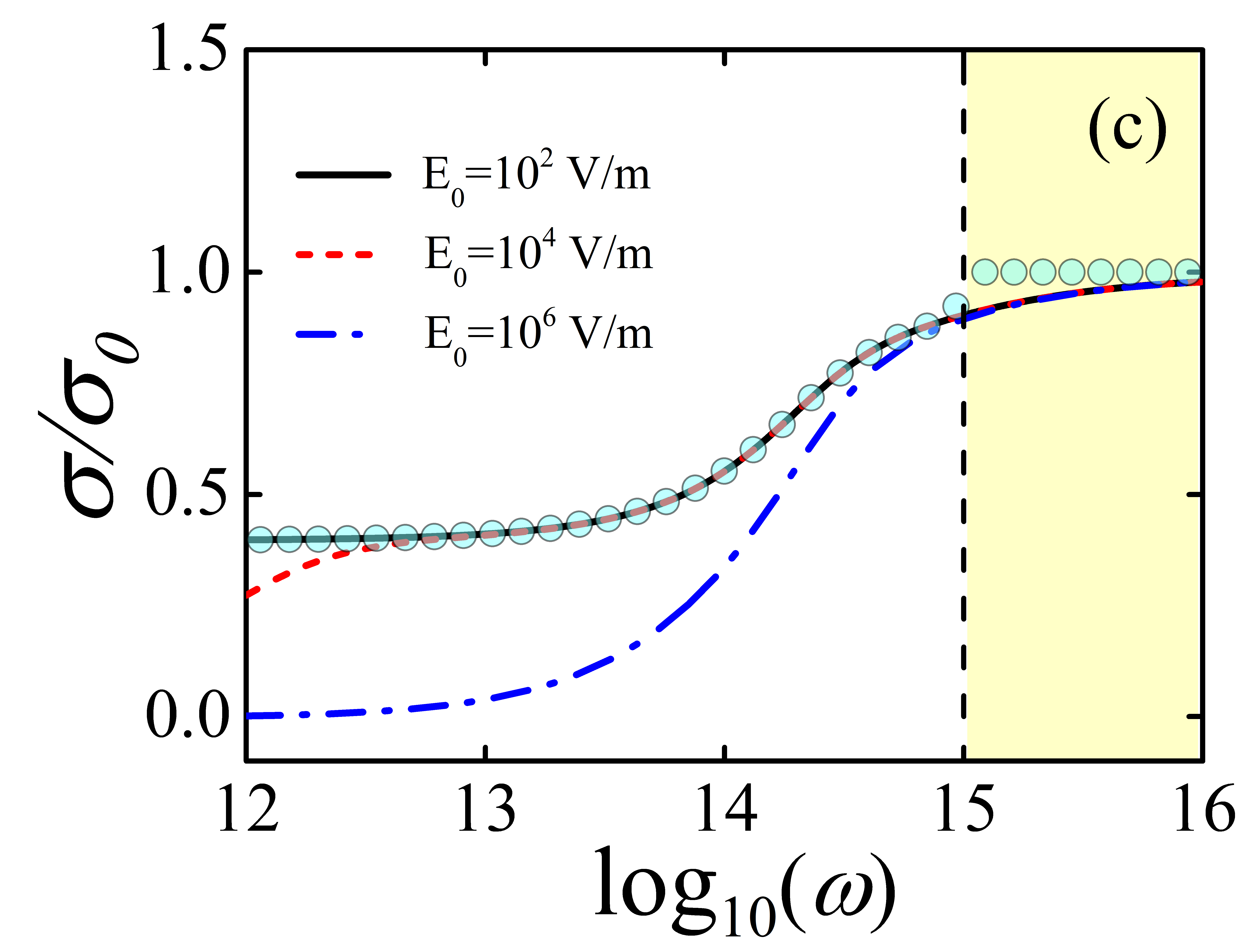}
        \includegraphics[width=4.2cm,height=3.5cm]{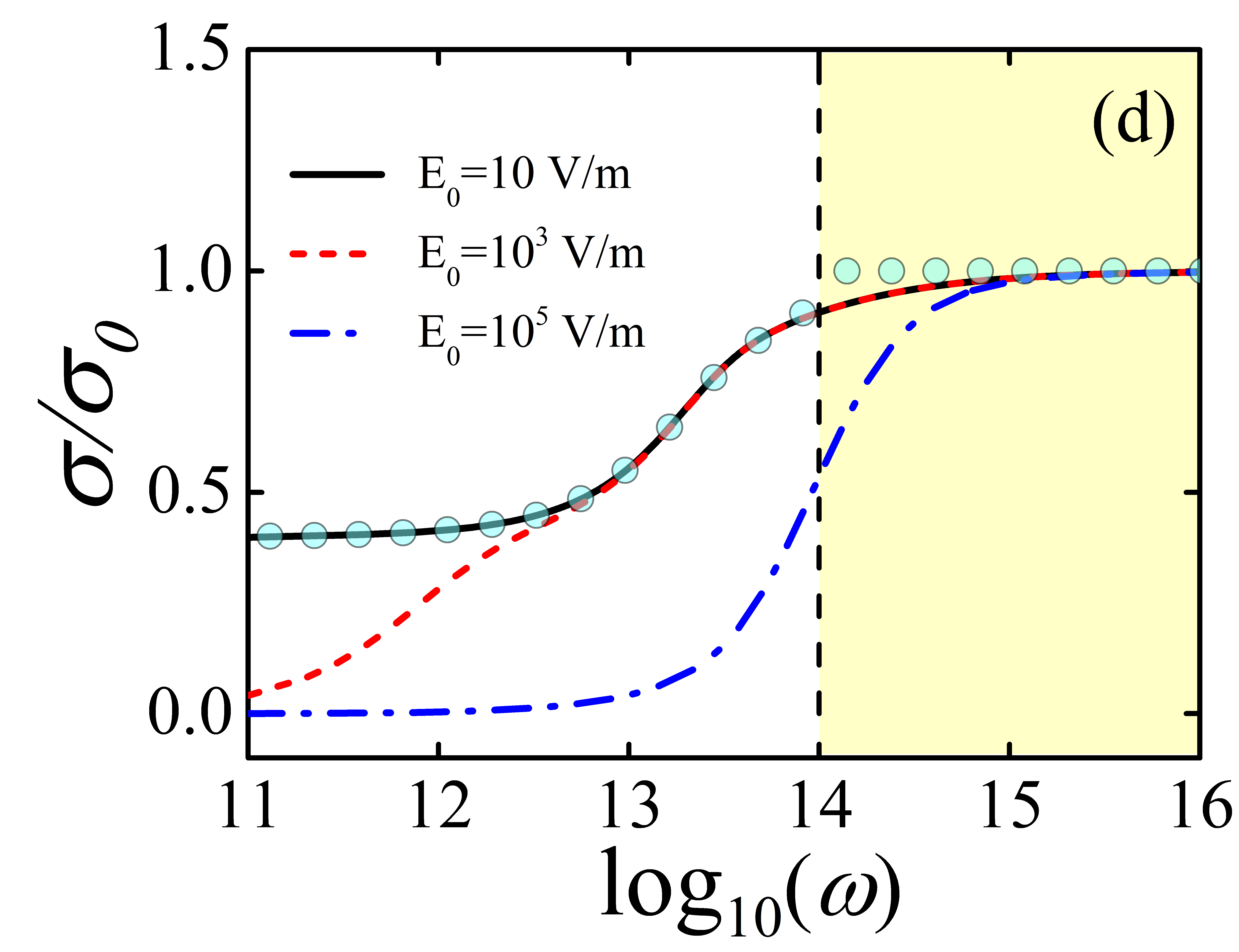}
	\caption {Plot of normalized optical conductivity $\sigma/\sigma_0$ ( $\sigma_0 = e^{2}/4\hbar$) of the pristine graphene as a function of frequency ($\omega$) and strength ($E_0$ ) of the external time varying electric field at
(a) 300 K and (b) 30 K temperature. For both the cases $\mu=0$ is considered. The white
line (dashed-dot) corresponding to the $\eta=1$, marks the linear and nonlinear domain of optical response of
the pristine graphene. The vertical white line (dash) at $\omega=\gamma_e$ is the boundary of clean
limit and dirty limit. These two white lines divide the whole plot into four regions which are linear clean (\emph{lc}), nonlinear clean (\emph{nc}), linear dirty (\emph{ld}) and nonlinear dirty (\emph{nd}). For plotting purpose Eq.(2) is used as it represents the most general case of the optical conductivity. Here,$v_F=10^{6}$ m/s is considered. The values of ($\gamma_p$,$\gamma_e$ ) for 300 K and 30 K are ($10^{12}$,$10^{14}$) Hz and ($10^{11}$,$10^{13}$) Hz, respectively. Plot of $\sigma/\sigma_0$ as a function of frequency ($\omega$) for three different electric fields at (c) 300 K and (d) 30 K temperature. The vertical dashed line at $\omega=\gamma_e$ marks the starting of the clean limit. The solid (cyan) circles represent Eq.(33) and Eq.(38) which agree well
 with the general numerical results (black solid line) in the linear clean limit (yellow region) and linear dirty limit (white region), respectively, proving the acceptability of the approximations in these two limits.}
\end{figure}
\subsubsection{Linear dirty regime:($\eta << 1, \frac{\gamma_e}{\omega }\geq 1$)}
   In this limit we can again approximate $\langle\Pi_\textbf{k}\rangle_{st} \approx\langle\Pi_\textbf{k}\rangle_{eq}$ in the lowest order of $\eta$. In contrary to the linear clean domain, we need to retain the Lorentzian part of the integrand in Eq.(30). Hence, the conductivity is given by,
   \begin{equation}
   \begin{aligned}
       \sigma_{xx}&=\dfrac{g_sg_\nu e^2 v^{2}_F}{\omega (2\pi)^{2}}\int\sin^{2}{\chi_\textbf{k}} \langle \Pi_\textbf{k} \rangle_{eq}\dfrac{\gamma_e}{\gamma^{2}_e+(\Delta_\textbf{k}-\omega)^{2}}\,d\textbf{k}.\\
       \end{aligned}
   \end{equation}
   Although one needs to compute this integral numerically for any arbitrary finite temperature, the closed form of it can be obtained at zero temperature. At zero temperature the function $g(\omega,\mu,T\to 0)=\Theta(\dfrac{\hbar\omega}{2}-|\mu|)$ which helps to rewrite Eq.(38)at zero temperature as
   \begin{equation}
       \begin{aligned}
          \sigma_{xx}&=\dfrac{e^2\gamma_e}{4\pi\omega}\int^{\dfrac{2\Lambda}{\hbar}}_{\dfrac{2|\mu|}{\hbar}}[\dfrac{\Delta_\textbf{k}}{\gamma^{2}_e+(\Delta_\textbf{k}-\omega)^{2}}-(\omega \to 0)]\,d\Delta_\textbf{k}\\
          &=\dfrac{e^2\gamma_e}{4\pi\omega}[\dfrac{1}{2}\ln{\dfrac{(\gamma^{2}_e+(\omega-\Delta_\textbf{k})^{2})}{\gamma^{2}_e+\Delta^{2}_\textbf{k}}}+\dfrac{\omega}{\gamma_e}\tan^{-1}{\dfrac{\Delta_\textbf{k}-\omega}{\gamma_e}}]^{\dfrac{2\Lambda}{\hbar}}_{\dfrac{2|\mu|}{\hbar}}.\\
       \end{aligned}
   \end{equation}
   Here, $\Lambda$ is nothing but the ultraviolet cut-off and it is usually considered as half of the bandwidth of graphene [ref].
   \subsubsection {Nonlinear dirty regime:($\eta > 1, \frac{\gamma_e}{\omega} \geq 1$)}
   This regime can be considered as most general domain where we can not apply any kind of approximation, and one needs to apply the  generalized expression of optical conductivity as given by Eq.(30). With the help of the full form of $\langle\Pi_\textbf{k}\rangle_{st}$ one may obtain
   \begin{equation}
       \begin{aligned}
           \sigma_{xx}&=\dfrac{-g_sg_v e^2 v^{2}_F}{\omega (2\pi)^{2}}\int\sin^{2}{\chi_\textbf{k}} \langle \Pi_\textbf{k} \rangle_{st}\dfrac{\gamma_e}{\gamma^{2}_e+(\Delta_\textbf{k}-\omega)^{2}}\,d\textbf{k}\\
         &=\dfrac{-g_sg_v e^2 v^{2}_F}{\omega (2\pi)^{2}}\int\dfrac{\sin^{2}{\chi_\textbf{k}} \langle\Pi_\textbf{k}\rangle_{eq}\gamma_e}{[(\Delta_\textbf{k}-\omega)^{2}+\gamma^2_e(1+\eta^2\sin^{2}{\chi_\textbf{k}})]}\,d\textbf{k}.\\
           \end{aligned}
   \end{equation}
         This equation can be evaluated at zero temperature as follows:
 \begin{eqnarray}
 \sigma_{xx} &=& \dfrac{e^2\gamma_e}{4\omega (\pi)^{2}}\int^{2\pi}_{0} \sin^{2}{\chi_\textbf{k}} d\chi_\textbf{k} \nonumber \\ &\times &\int^{\dfrac{2\Lambda}{\hbar}}_{\dfrac{2|\mu|}{\hbar}}(\dfrac{\Delta_\textbf{k}}{[(\Delta_\textbf{k}-\omega)^{2}+\gamma^2_e(1+\eta^2\sin^{2}{\chi_\textbf{k}})]}-(\omega \rightarrow  0))d\Delta_\textbf{k} \nonumber\\
  &=&\dfrac{e^2}{4(\pi)^{2}}\int^{2\pi}_{0}\sin^{2}{\chi_\textbf{k} d\chi_\textbf{k}}[f_1(\omega,\dfrac{2\Lambda}{\hbar})-f_1 (\omega,\dfrac{2|\mu|}{\hbar})], \nonumber \\
 \end{eqnarray}
    where
    \begin{equation}
        \begin{aligned}
            f_1(\omega,x)=\dfrac{\gamma_e}{\gamma^{2}_{1}}\tan^{-1}{\dfrac{\Delta_\textbf{k}-\omega}{\gamma^{2}_{1}}}+\dfrac{\gamma_e}{2\omega}\ln{\dfrac{(\gamma^{2}_{1}+(\omega-\Delta_\textbf{k})^{2})}{\gamma^{2}_{1}+\Delta^{2}_\textbf{k}}},
        \end{aligned}
    \end{equation}
and $\gamma^{2}_{1}=\gamma_e\sqrt{1+\eta^2\sin^{2}{\chi_\textbf{k}}}$.\\
\indent
It is pertinent to mention here that the expression of conductivity (at zero temperature) in all regimes exactly matches with the previously obtained results [19]. Let us explain Figure 3. The colour plot of the optical conductivity of pristine graphene with various $E_0$ and different $\omega$ is shown in Figure (3a) and Figure (3b) at 300 K and 30 K temperatures, respectively. Here, the lines $\eta=1$  and $\omega=\gamma_e$ divide the entire colour plot into four regions : 1. linear clean (\emph{lc}), 2. nonlinear clean (\emph{nc}), 3. linear dirty (\emph{ld}), and 4. nonlinear dirty (\emph{nd}). It can be observed from the Figure (3) that the starting frequency of clean limit is red-shifted at low temperature (30 K) due to the temperature dependency of $\gamma_e$  and $\gamma_p$. As a matter of fact one may observe that the position of the line corresponding to the Mischenko parameter $\eta=1$ changes with the change of temperatures. In the ‘nonlinear dirty’ region with high $E_0$, the optical conductivity does not increase with the increase of the field strength $E_0$ , rather it shows a saturation behavior. At high enough incident electric field the absorption coefficient decreases significantly due to the depletion of carriers in valence band and results in saturation effect of the optical conductivity.  This typical behaviour of optical conductivity is also shown in Figure (3c) and Figure (3d) where the conductivity is almost zero in the dirty limit for $E_0=10^6$ V/m and $E_0=10^5$ V/m , respectively. In the clean limit, when $\omega>>\gamma_e$, the optical conductivity of pristine graphene approaches its universal value ($\sigma_0$) for both high temperature (300 K) and low temperature (30 K)  as shown in Figure (3c) and Figure (3d), respectively.
   \section{Gapped Graphene}
   We now analyze the nonlinear optical conductivity for the case of a gap graphene that can be obtained by introducing a gap 2$\Delta$ in the band structure of graphene. For example, gapped can be introduced in graphene when it is epitaxially grown on any substrate [29].  Again, we follow the same kind of spin-Boson type model and quantum master equation method to analyze the nonlinear optical conductivity of gapped graphene. If a gap 2$\Delta$ is created in the graphene band structure, the Hamiltonian of the subsystem is modified as follows
  \begin{equation}
\begin{aligned}
   H_S&=v_F(\sigma\cdot \textbf{k})+\Delta\sigma_z.\\
    \end{aligned}
\end{equation}
The eigenvalues of  $H_S$ are given by $\pm \sqrt{(\Delta^{2} +(v_F k)^{2})}$. To simplify it further we may consider $\Delta=a_0\cos{\theta}$ and $v_Fk=a_0\sin{\theta}$, so that the eigenvalues become $\pm a_0$ and the eigenfunctions of modified $H_S$ are
\begin{equation}
   |c_k\rangle =
   \begin{pmatrix}
\exp({-i\chi_{\bf{k}}})\cos{\dfrac{\theta}{2}}\\[1em]
\sin{\dfrac{\theta}{2}}
\end{pmatrix},
|v_k\rangle =
   \begin{pmatrix}
-(\exp{-i\chi_{\bf{k}}})\sin{\dfrac{\theta}{2}}\\[1em]
\cos{\dfrac{\theta}{2}}
\end{pmatrix},
\end{equation}
where $\chi_{\bf{k}}$ represents the angle formed by the $\textbf{k}$ vector with $x$ axis. Let us introduce the total Hamiltonian of the gapped graphene as follows :
\begin{equation}
 H(t)=H_S+H_\omega(t)+H_{SB}+H_B,
\end{equation}
where, $H_B$ and $H_\omega(t)$ have the same form as that of Eq.(4) and the second term of Eq. (5), respectively. As mentioned earlier our system Hamiltonian is simplified as :
\begin{equation}
H_S=a_0\Pi_\textbf{k},
\end{equation}
while, the interaction term becomes
\begin{equation}
H_{SB}=\Pi_\textbf{k}X_e+(Y^+_\textbf{k}+Y^-_\textbf{k})X_p.
\end{equation}
On the other hand, the `effective' ac term for gapped graphene can be written as :
\begin{equation}
    \begin{aligned}
        H^{eff}_\omega (t)
        &=[-\exp{(-i(\omega - B_{\textbf{k}})t})\Omega_{\textbf{k}} Y^+_{\textbf{k}}\\
        &-\exp{(i(\omega - B_{\textbf{k}})t})\Omega^+_{\textbf{k}}Y^-_{\textbf{k}}]/2,
        \end{aligned}
\end{equation}\\
where
\begin{equation}
    \begin{aligned}
       \Omega_\textbf{k}&=(eEv_F/\omega) (\sin{\chi_\textbf{k}}-i\cos{\theta}\cos{\chi_\textbf{k}})\\
       \Omega^{+}_\textbf{k}&=(eEv_F/\omega)(\sin{\chi_\textbf{k}}+i\cos{\theta}\cos{\chi_\textbf{k}}) \\
    B_{\textbf{k}}&=2\sqrt{\Delta^2+(v_Fk)^2},
    \end{aligned}
\end{equation}
and,
\begin{equation}
    \begin{aligned}
        \begin{split}
        H_{SB}(t) &=\Pi_\textbf{k}X_e(t)+[\exp({iB_{\textbf{k}} t})Y^{+}_\textbf{k}+\exp({-iB_{\textbf{k}} t})Y^{-}_\textbf{k}]X_p(t).
        \end{split}
    \end{aligned}
\end{equation}
With the help of similar approach as that of Sec.(IIB), we can write the equations which govern the dynamics of $\langle\Pi_{\bf{k}}(t)\rangle$, and $\langle Y_{\bf{k}}^{+}(t)\rangle$  for gapped graphene as follows :
\begin{equation}
    \begin{aligned}
      \begin{split}
          \dfrac{d}{dt}\langle\Pi_\textbf{k} (t)\rangle&=[i\Omega_\textbf{k}\langle Y^+_\textbf{k}\rangle\exp(-i\omega t)-i\Omega^+_\textbf{k}\langle Y^{-}_\textbf{k}\rangle\exp(i\omega t)] \nonumber \\
          &-\gamma_p[\langle\Pi_\textbf{k} (t)\rangle-\langle\Pi_\textbf{k}\rangle_{eq}],
      \end{split}
    \end{aligned}
\end{equation}
and
\begin{equation}
\dfrac{d}{dt}\langle Y^{+}_\textbf{k} (t)\rangle=-\dfrac{i\Omega^{+}_{\textbf{k}}}{2}\langle\Pi_\textbf{k}\rangle \exp(i\omega t)-(\gamma_e-iB_{\textbf{k}})\langle Y^{+}_\textbf{k}\rangle.
\end{equation}
Again we can introduce the momentum resolved current density of the gapped graphene along the applied electric field direction as
\begin{eqnarray}
   j_{\textbf{k}x} &=& ev_F[\sin{\theta}\cos{\chi_\textbf{k}}\langle \Pi_\textbf{k}\rangle+(\cos{\theta}\cos{\chi_\textbf{k}}+i\sin{\chi_\textbf{k}})\langle Y^{+}_\textbf{k}\rangle\nonumber \\&+&(\cos{\theta}\cos{\chi_\textbf{k}}-i\sin{\chi_\textbf{k}})\langle Y^{-}_\textbf{k}\rangle].
\end{eqnarray}
In the steady state, the momentum dependent current density for gapped graphene is
\begin{eqnarray}
&& j_{\textbf{k}x} (t)_{st} = ev_F[\sin{\theta}\cos{\chi_\textbf{k}}\langle \Pi_\textbf{k}(t)\rangle_{st}+\cos{\theta}\cos{\chi_\textbf{k}}\nonumber \\
&\times&[\langle Y^+_\textbf{k}(t)\rangle_{st}+\langle Y^-_\textbf{k}(t)\rangle_{st}]
    +i\sin{\chi_\textbf{k}}[\langle Y^+_\textbf{k}(t)\rangle_{st}-\langle Y^-_\textbf{k}(t)\rangle_{st}]].\nonumber \\
\end{eqnarray}
We can write the conductivity for gapped graphene as follows :
\begin{eqnarray}
  \sigma_{xx}&=\dfrac{-g_sg_\nu e^2 v^{2}_F}{\omega (2\pi)^{2}}\int(\sin^{2}(\chi)+\cos^2{\theta}\cos^2({\chi}) \langle \Pi_\textbf{k} \rangle_{st}\nonumber \\
&\times\dfrac{\gamma_e}{(B_\textbf{k}-\omega)^{2}+\gamma^{2}_e} d\textbf{k},
\end{eqnarray}
where
\begin{equation}
    \langle\Pi_\textbf{k}\rangle_{st}=\langle\Pi_\textbf{k}\rangle_{eq}[1+\dfrac{\eta^2 \gamma^{2}_e (\sin^2{\chi_\textbf{k}}+\cos^2{\chi_\textbf{k}}\cos^2{\theta})}{\gamma^{2}_e+(B_{\textbf{k}}-\omega)^{2}}]^{-1}.
\end{equation}
\subsection{Nonlinear optical conductivity}
 We want to  study the optical conductivity of a gapped graphene for different regimes similar to the pristine graphene, as discussed earlier. Here, for calculating the conductivity, assumptions are made similar to the Sec.(IID).
 \subsubsection{Linear clean limit: ($\eta << 1$, ${\gamma_e)}{\omega} << 1$)}
 In this limit, the conductivity can be written as
    \begin{equation}
        \begin{aligned}
          \sigma_{xx}
          &=\dfrac{e^2 g_sg_\nu  g(\omega,\alpha,T)}{16}[\dfrac{4\Delta^2}{\omega^2}+1],
        \end{aligned}
    \end{equation}
  where $\alpha = max[\mu,\Delta]$ [16]. The function $g(\omega,\alpha,T)$ is given by
    \begin{equation}
   g(\omega,\alpha,T)=\dfrac{1}{2}[\tanh{\dfrac{\omega+2\alpha}{4k_{B}T}}+\tanh{\dfrac{\omega-2\alpha}{4k_{B}T}}].
    \end{equation}
    In the limiting case of $T\rightarrow 0$, $g(x)\rightarrow \Theta(x)$ and
    \begin{equation}
    \sigma_x =\dfrac{e^2}{4}[\dfrac{4\Delta^2}{\omega^2}+1]\Theta(\dfrac{\omega}{2}-\alpha).
    \end{equation}
   One may observe that the effect of chemical potential is unimportant if it lies inside the gap ($\mu < \Delta$). One can recover the normal graphene results in the limit $\Delta \rightarrow 0 $.
 \begin{figure}[h]
	\centering
	\includegraphics[width=4.2cm,height=3.6cm]{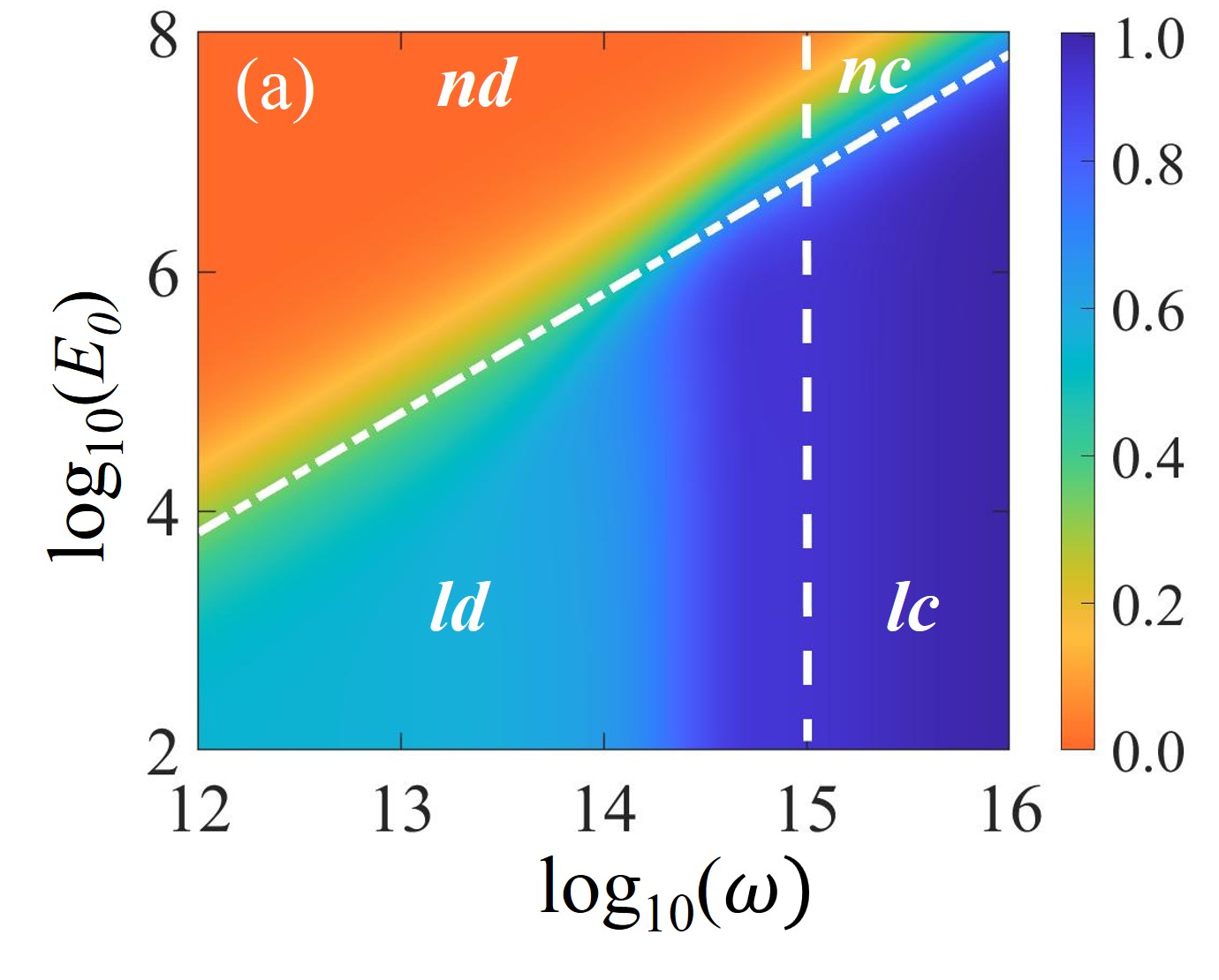}
        \includegraphics[width=4.3cm,height=3.6cm]{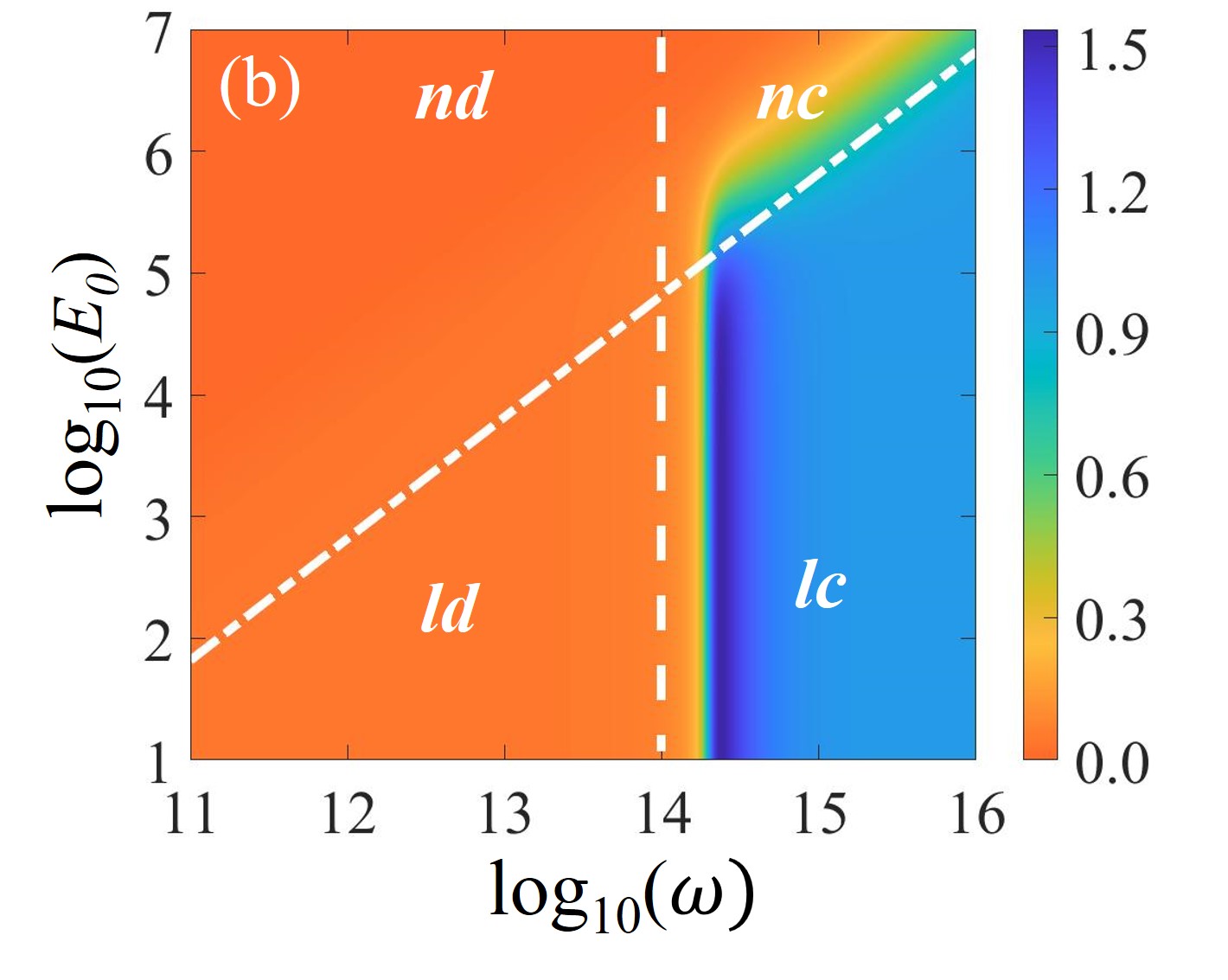}
        \includegraphics[width=4.2cm,height=3.6cm]{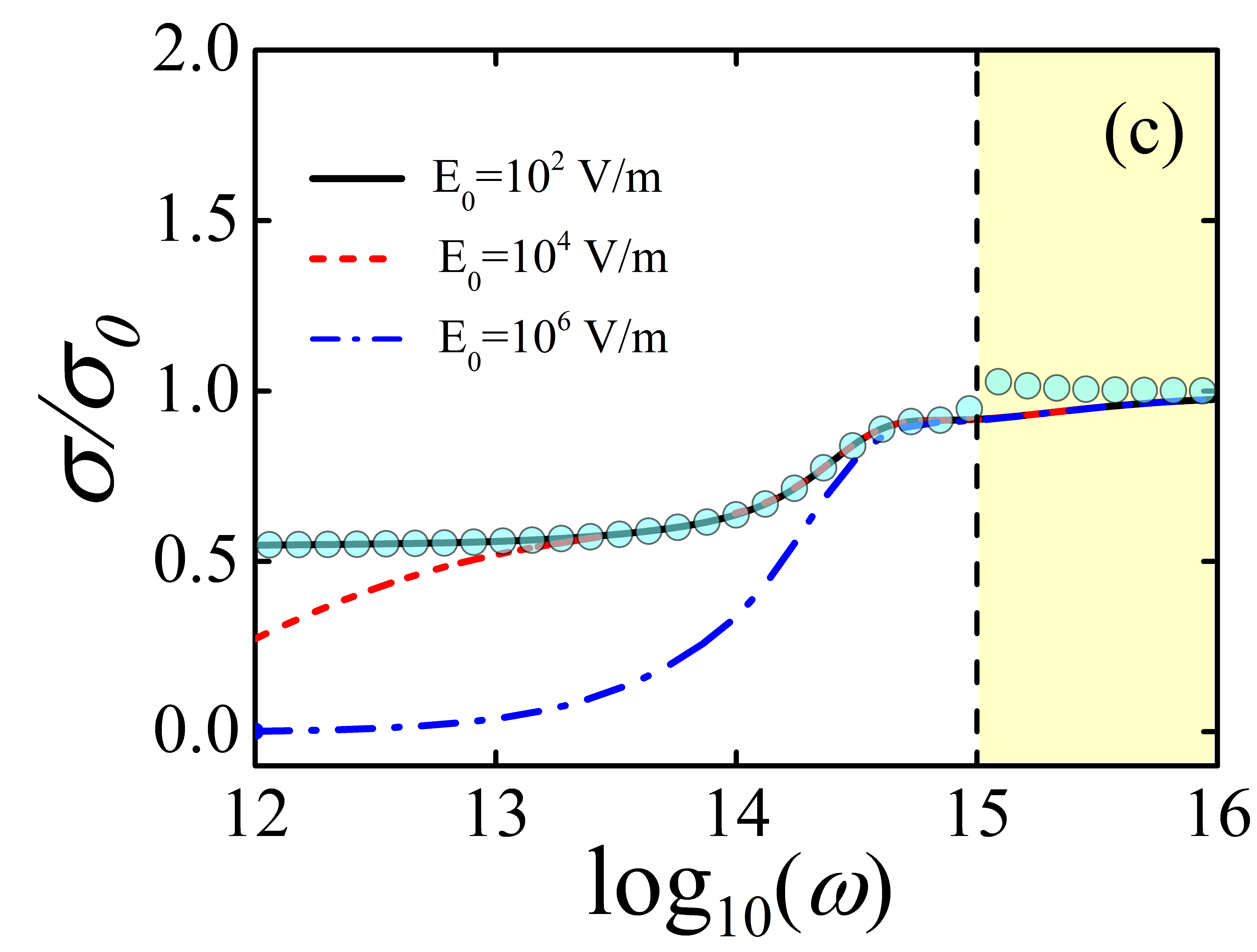}
        \includegraphics[width=4.2cm,height=3.6cm]{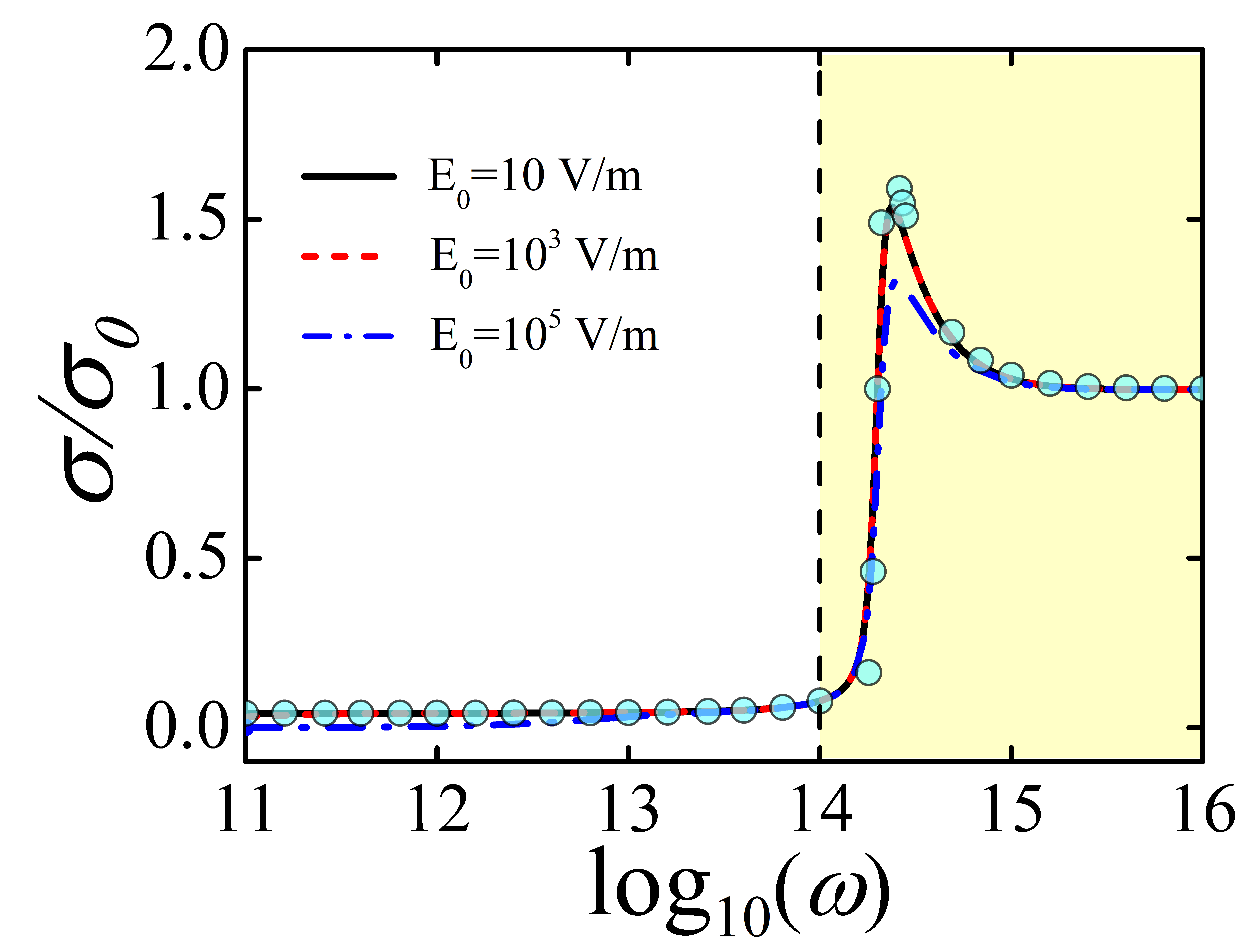}
	\caption {Plot of normalized optical conductivity $\sigma/\sigma_0$ ( $\sigma_0 = e^{2}/4\hbar$) of the gapped graphene as a function of frequency ($\omega$) and strength ($E_0$ ) of the external time varying electric field at
(a) 300 K and (b) 30 K temperature. For both the cases $\mu=0$ and $\Delta =0.065 $ eV are considered. The white
line (dashed-dot) corresponding to the $\eta=1$, marks the linear and nonlinear domain of optical response of
the pristine graphene. The vertical white line (dash) at $\omega=\gamma_e$ is the boundary of clean
limit and dirty limit. These two white lines divide the whole plot into four regions which are
linear clean (\emph{lc}), nonlinear clean (\emph{nc}), linear dirty (\emph{ld}) and nonlinear dirty (\emph{nd}). For plotting  purpose Eq.(54) is used, as it represents the most general case of the optical conductivity for gapped graphene. Here, $v_F=10^{6}$ m/s is considered. The values of ($\gamma_p$,$\gamma_e$ ) for 300 K and 30 K are ($10^{12}$,$10^{14}$) Hz and ($10^{11}$,$10^{13}$) Hz respectively. Plot of $\sigma/\sigma_0$ as a function of frequency ($\omega$) for three different electric fields at (c) 300 K and (d) 30 K temperature. The vertical dashed line at $\omega=\gamma_e$  marks the starting of the clean limit. The blue circles representing the Eq. (56) and Eq. (61) which agree with the general numerical results (black solid line) in the linear dirty limit (white region) and linear clean limit (yellow region) respectively, proving the acceptability of the approximations in these two limits.}
\end{figure}
   \subsubsection{Nonlinear, clean limit:($\eta\geq 1, \frac{\gamma_e}{\omega} << 1$)}
   In this regime, the Lorentzian can be approximated by a delta function, and one obtains
   \begin{equation}
         \sigma_{xx}
         =\dfrac{e^2g(\omega,\alpha,T)}{2\eta^2}[1-\dfrac{1}{\sqrt{1+\eta^2}}(1+\dfrac{4\Delta^2 \eta^2}{\omega^2})^{-\dfrac{1}{2}}].
   \end{equation}

\subsubsection{Linear dirty limit:($\eta << 1$,$\frac{\gamma_e}{\omega} \geq 1$)}
   In this limit the conductivity is given as,
   \begin{eqnarray}
          \sigma_{xx} &=&\dfrac{ -g_sg_\nu e^2 v^{2}_F}{\omega (2\pi)^{2}}\int(\cos^2{\theta}\cos^2{\chi}+\sin^2{\chi} )\langle \Pi_\textbf{k} \rangle_{eq}\nonumber \\
          &\times& \dfrac{\gamma_e}{\gamma^{2}_e+(B_{\textbf{k}}-\omega)^{2}}\,d\textbf{k},
   \end{eqnarray}
   Now one can obtain closed form expression for zero temperature, and it is given by
   \begin{equation}
       \begin{aligned}
          \sigma_{xx}&=\dfrac{ e^2 \gamma_e}{4\pi\omega} \int^{\dfrac{2\Lambda}{\hbar}}_{\dfrac{2|\mu|}{\hbar}}[\dfrac{4\Delta^2 + B_{\textbf{k}}^2}{B_{\textbf{k}}(\gamma^{2}_e+(B_{\textbf{k}} -\omega)^{2})}-(\omega \to 0)]\,dB_{\textbf{k}}\\
          &=\dfrac{e^2\gamma_e}{4\pi\omega}[f_2(\omega,\dfrac{2\Lambda}{\hbar})-f_2 (\omega,\dfrac{2|\mu|}{\hbar})],\\
       \end{aligned}
   \end{equation}
   where
   \begin{eqnarray}
         &&f_2(w,x) =(1+y)\tan^{-1}{(\dfrac{x-\omega}{\gamma_e})}-\dfrac{\gamma^2_e-4\hbar^{-2}\Delta^2}{2\gamma_e\omega} \nonumber \\
         &\times&\ln{[x^2+\gamma^2_e]}
         +\dfrac{\gamma_e(1-y)}{2\omega}\ln{[(x-\omega)^2+\gamma^2_e]}-\dfrac{\omega y}{\gamma_e}\ln{x},\nonumber \\
   \end{eqnarray}
   and $y=\dfrac{4\hbar^{-2}\Delta^2}{\omega^2 + \gamma^2_e}$.
    \subsubsection{Nonlinear,dirty regime:($\eta > 1, {\gamma_e}/{\omega} \geq 1$)}
   In this limit, we have to use the most generalized expression as given by Eq.(54). Thus, the conductivity has the form,
   \begin{eqnarray}
           &&\sigma_{xx} =\dfrac{-g_sg_\nu e^2 v^{2}_F}{\omega (2\pi)^{2}}\int A(\theta,\chi) \langle \Pi_\textbf{k} \rangle_{st}\dfrac{\gamma_e}{\gamma^{2}_e+(B_{\textbf{k}}-\omega)^{2}} d\textbf{k} \nonumber\\
         &=& \dfrac{-g_sg_\nu e^2 v^{2}_F}{\omega (2\pi)^{2}}\int\dfrac{A(\theta,\chi) \langle\Pi_\textbf{k}\rangle_{eq}\gamma_e}{[(B_{\textbf{k}}-\omega)^{2}+\gamma^2_e(1+\eta^2A(\theta,\chi))]}d\textbf{k}, \nonumber\\
   \end{eqnarray}
 where $A(\theta,\chi)=(\sin^{2}{\chi+\cos^2{\theta}\cos^2{\chi}})$. It can be noted that our results for the optical conductivity (lc, ld, nc, nd regimes) of gapped graphene are in good agreement with the previous study of A. Singh et. al [19]. \\
 Now we can do numerical evaluation of optical conductivity. As one may observe, the Figure (4a) and Figure (4b) demonstrate the plots of the optical conductivity of gapped graphene, which are also divided into the four regimes by the two lines $\eta=1$  and $\omega=\gamma_e$ as we previously mentioned for pristine graphene. These two lines differ for both graphs depending on temperature. On the other hand,  Figure (4c) and  Figure (4d) show the variation of optical conductivity with the frequency $\omega$ for different incident electric fields at 300 K and 30 K temperatures, respectively. Here we can see an intriguing phenomenon for 30 K, where throughout the dirty region (Figure (4b)) the optical conductivity assumes a saturation value near zero irrespective of applied electric field strength. Figure (4d) clearly shows the same kind of phenomenon where the conductivity is almost zero up to $10^{14}$ Hz for all three different incident electric fields. Then there is a sudden jump in conductivity after a specific frequency inferring the major role of the band gap behind the process. Further to investigate the role of bandgap in the low-temperature optical conductivity process of the gapped graphene, optical conductivity is calculated for different values of $\Delta$. Figures (5a) and (5b) illustrate the frequency dependency of the optical conductivity of the gapped graphene with different band gaps (2$\Delta$) at  300 K and 30 K temperatures. For 300 K temperature, the optical conductivity varies continuously with frequency for all the band gap values (Figure (5a)). But the case becomes more interesting for low temperature (30 K), where,  at a prominent band gap value, the conductivity changes abruptly after a particular frequency as shown in Figure 5b. This abrupt change in conductivity does not occur for small values of the bandgap rather the conductivity changes gradually. At lower temperatures (30 K), following the Fermi distribution, the conduction band (upper Dirac Cone) lacks carriers. For pristine graphene, the gap-less band structure helps to absorb a broad spectrum of the incident light in the linear dirty region, hence a non-zero absorption coefficient or non-zero optical conductivity can be observed even at a low temperature, as reflected in Figure (3b) and Figure (3d).  But for gapped graphene, carriers only can absorb the optical pulse when a minimum 2$\Delta$ amount of energy is supplied by the external driving field of the optical pulse to the carriers for the required bandgap transition. At any incident frequency lower than the bandgap frequency ($\hbar\omega= 2\Delta$), the optical conductivity becomes almost zero irrespective of electric field strength ( Figure (4b) and Figure (4d)). When the frequency of optical pulse reaches a specific frequency equivalent to bandgap, it is instantaneously absorbed by the carriers. Consequently, the optical conductivity of the gapped graphene takes a sudden jump at that particular frequency (Figure (4d) and Figure (5b)).
\begin{figure}[h]
	\centering
	\includegraphics[width=4.2cm,height=3.5cm]{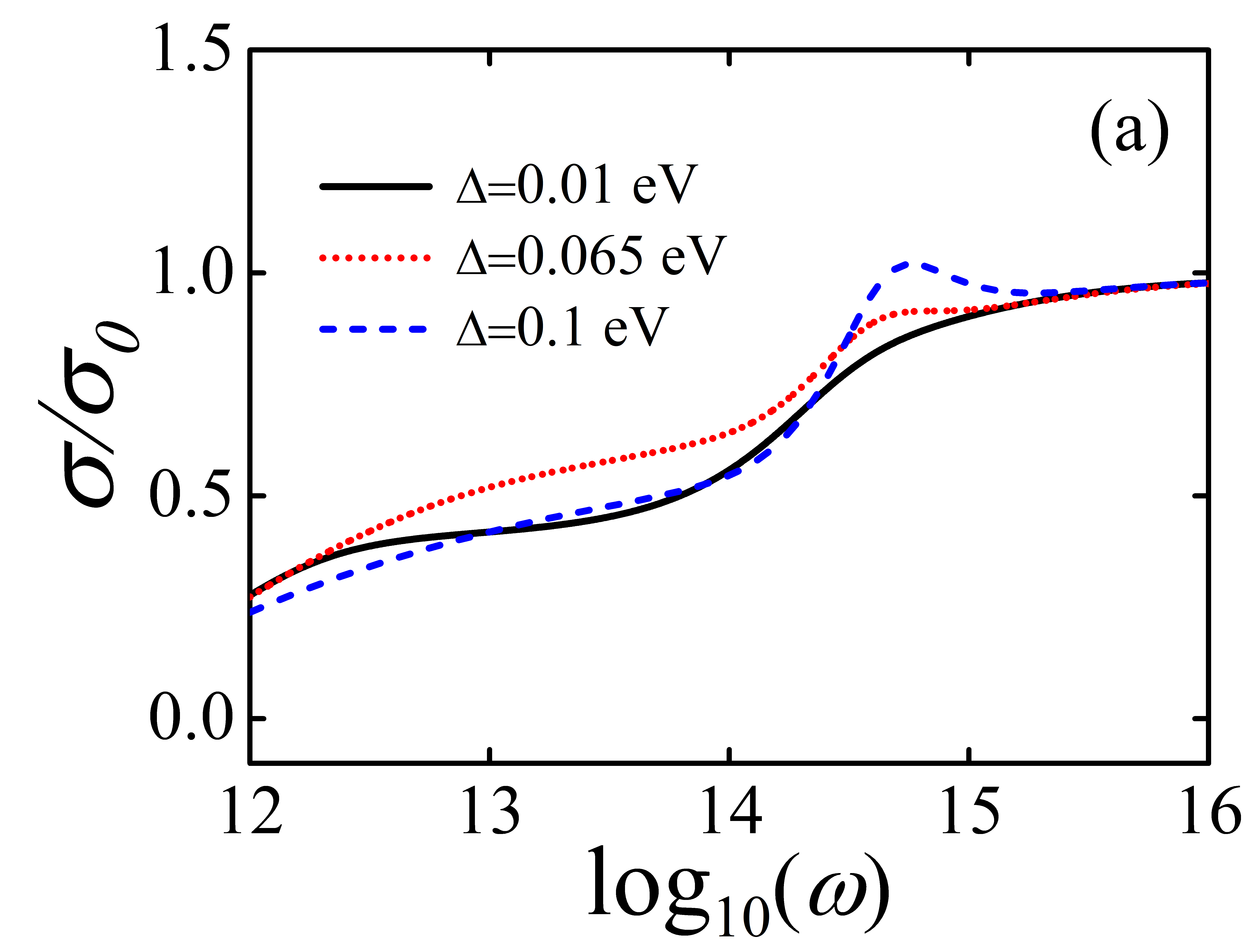}
        \includegraphics[width=4.2cm,height=3.5cm]{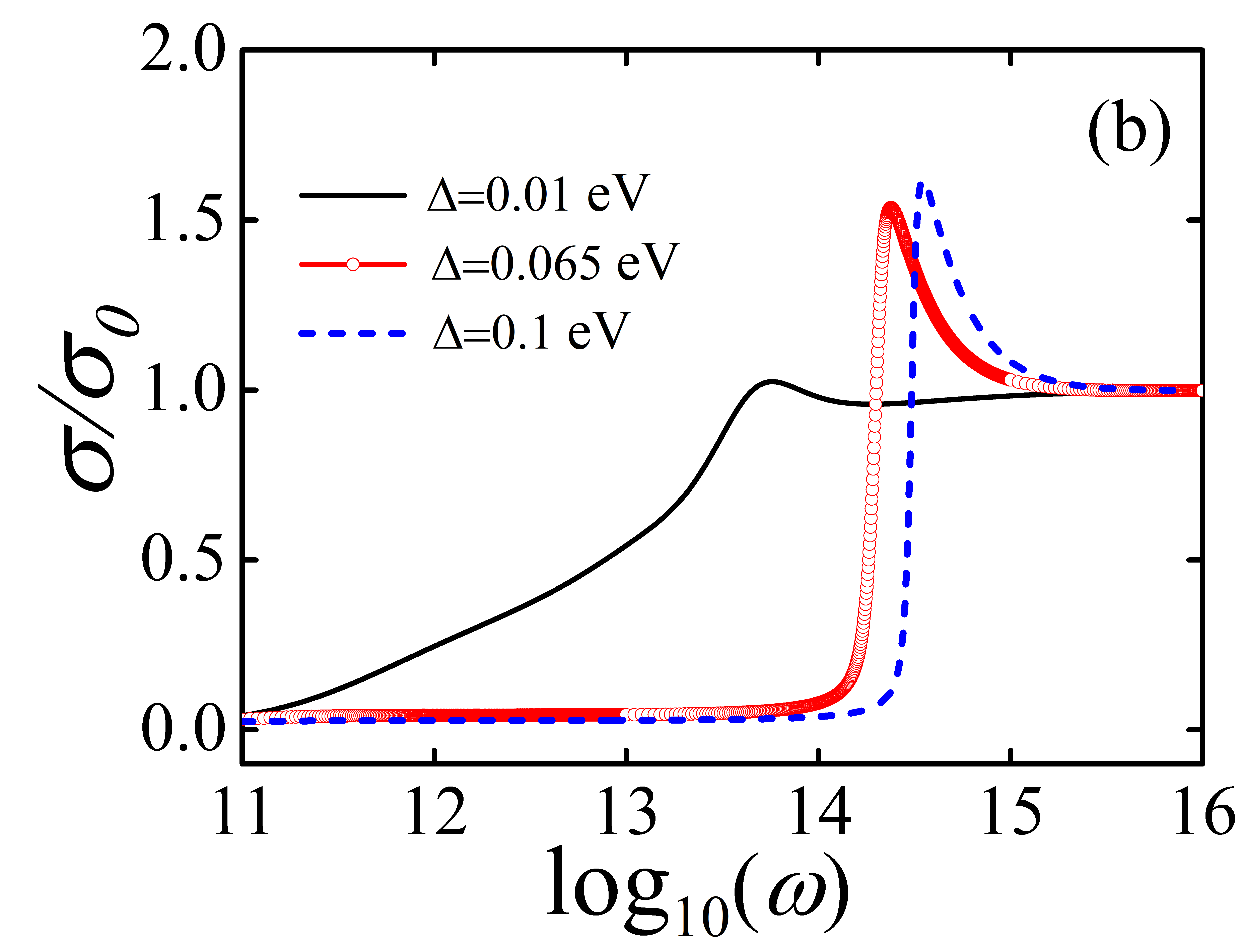}
	\caption { Plot of normalized optical conductivity $\sigma/\sigma_0$ ( $\sigma_0 = e^{2}/4\hbar$) of the gapped graphene as a function of frequency ($\omega$) for different values of $\Delta$ at (a) 300 K and (b) 30 K temperature.}
	\label{EEDl}
\end{figure}\\
\section{Conclusion}
Our starting point in this paper has been the microscopic spin-boson approach to the nonlinear optical conductivity problem in graphene [20] which is a generalization of the phenomenological rate theory calculation [18, 19]. Like Ref. [18] and [20], we work in the rotating wave approximation in which terms off-resonant with the applied oscillatory field are ignored. However, we have gone beyond in making elaborate analyses of the relaxation rates mediated by the surrounding phonons and electrons of the graphene system. For this, separate spectral densities for the phonon and the electron baths have been incorporated in the analysis. Of special renewed interest has been the transient non-Markov regimes wherein strong quantum effects are observed that can be probed by presently available ultrafast spectroscopy techniques. We have also carefully delineated the Markov and non-Markov domains and transitions between them. One other feature that we have investigated here in detail, which was not covered in our earlier work [20], is the case of gapped graphene that brings-in different attributes [19]. A detailed analysis reveals characteristic properties of the graphene system in different regions of the Mischenko parameter values.\\
\indent
Although our interest in this paper has been restricted to the quantum solid of graphene, the methodology employed here is of relevance to general theoretical methods for dissipative behaviour of open quantum systems that belong to Non-equilibrium Statistical Mechanics. The resultant treatment sheds further light on the phenomenological approach adopted in Ref. [19] in terms of our microscopic method in which the bath parameters such as the cutoff frequency and the temperature appear explicitly. However, the mathematical formalism is brought to the domain of experiments on relaxation studies in graphene (such as Ref. [27] and [28]).
\section*{Acknowledgements}
B.G is supported by INSPIRE, DST, Government of India (IF200292). SD is grateful to the Indian National Science Academy for support through their Honorary Scientist scheme. M.B. is supported by the Department of Science and Technology (DST), Government of India under the Core grant (Project No. CRG/2020//001768) and MATRICS grant (Project no. MTR/2021/000566).

\appendix

\section{Spin-lattice relaxation time at High-T }
\begin{equation}
    \begin{aligned}
    \gamma_p&=4\int^{t}_0 cos{(\Delta_{\textbf{k}}\tau)}d\tau\int^{\infty}_0J_p(\omega)\dfrac{2k_B T}{\hbar\omega}\cos{(\omega\tau)} d\omega\\
    &=\dfrac{16\alpha_{e}k_B T\omega_{cp}}{\hbar}[\int^{t}_0 d\tau \cos{(\Delta_{\textbf{k}}\tau)}\dfrac{(1-3\omega^{2}_{cp}\tau^{2})}{(1+\omega^{2}_{cp}\tau^{2})^2}],\\
    \end{aligned}
\end{equation}
where the integrand part becomes,
\begin{equation}
    \begin{aligned}
    &\int^{t}_0 d\tau \cos{(\Delta_{\textbf{k}}\tau)}\dfrac{(1-3\omega^{2}_{cp}\tau^{2})}{(1+\omega^{2}_{cp}\tau^{2})^2}\\
    &=\dfrac{\Delta^2_{\textbf{k}}}{\omega^3_{cp}}\Bigg[i \cosh{\Big (\dfrac{\Delta_{\textbf{k}}}{\omega_{cp}}}\Big)\Big[ Ci(\Delta_{\textbf{k}} t+\dfrac{i\Delta_{\textbf{k}}}{{\omega_{cp}}})\\
    &-Ci(\Delta_{\textbf{k}} t-\dfrac{i\Delta_{\textbf{k}}}{{\omega_{cp}}})-Ci(\dfrac{i\Delta_{\textbf{k}}}{{\omega_{cp}}})+Ci(\dfrac{-i\Delta_{\textbf{k}}}{{\omega_{cp}}})-i\pi\Big]\\
     &+\sinh{\Big (\dfrac{\Delta_{\textbf{k}}}{\omega_{cp}}}\Big)\Big[ Si(-\Delta_{\textbf{k}} t+\dfrac{i\Delta_{\textbf{k}}}{{\omega_{cp}}})-Si(\Delta_{\textbf{k}} t+\dfrac{i\Delta_{\textbf{k}}}{{\omega_{cp}}})\Bigg]\\
    &+\dfrac{t\cos{(\Delta_{\textbf{k}} t)}}{(\omega^2_{cp}t^2+1)^2}-\dfrac{\Delta_{\textbf{k}}\sin{(\Delta_{\textbf{k}} t)}}{2(\omega^4_{cp}t^2+\omega^2_{cp})}.
    \end{aligned}
\end{equation}

\section{Spin-lattice relaxation time at low-T }
\begin{equation}
    \begin{aligned}
\gamma_p(t)&=24\alpha_p\omega_{cp}\Big[\int_{0}^{\omega_{cp}t}dx\frac{[1-6x^2+x^4]}{[1+x^2]^4}]\cos(bx)  \\
&+2 \int_{0}^{\omega_{cp}t}dx \frac{[(1+a_p)^4-6(1+a_p)^2x^2+x^4]}{[(1+a_p)^2+x^2]^4}\cos(bx) \Big].\\
    \end{aligned}
\end{equation}
The first integrand part is,
\begin{equation}
    \begin{aligned}
    &\int_{0}^{\omega_{cp}t}dx\frac{[1-6x^2+x^4]}{[1+x^2]^4}]\cos(bx)\\
        &=\dfrac{1}{12}\Bigg[ b^3(i\sinh{(b
)}Ci(-b(l-i))-i\sinh{(b)}Ci(b(l+i))\\
&+\cosh{(b)}Si(b(x+i))-Si(-b(l-i)))\\
&+\dfrac{2l(b^2(l^{2}+1)^2 - 2l^2 +6)\cos{(bl)}}{(l^{2}+1)^3}+\dfrac{2b(l^2-1)\sin{(bl)}}{(l^{2}+1)^2}\Bigg],
    \end{aligned}
\end{equation}
and the second integrand becomes,
\begin{equation}
    \begin{aligned}
        &\int_{0}^{\omega_{cp}t}dx \frac{[(1+a_p)^4-6(1+a_p)^2x^2+x^4]}{[(1+a_p)^2+x^2]^4}\cos(bx)\\
        &=\dfrac{1}{6}\Bigg[\dfrac{b(-a^2_{p}-2a_{p}+l^2 _1)\sin{(bl)}}{(a^2_{p}+2a_{p}+l^2+1)^2}\\
        &+\dfrac{l\cos{(bl)}(a^4_{p}b^2 +4a^3_{p}b^2+2a^2_{p}(b^2 (l^2+3)+3)}{(a^2_{p}+2a_{p}+l^2+1)^3}\\
        &+\dfrac{(4a_p (b^2 (l^2+1)+3)+b^2(l^2 +1)-2l^2 +6))l\cos{(bl)}}{(a^2_{p}+2a_{p}+l^2+1)^3}\\
        &+\dfrac{b^3}{2}(i\sinh{((a_p +1)b)}(Ci(ib(a_p+il+1))-Ci(ib(a_p-il+1)))\\
        &+\cosh{((a_p +1)b)(Si(b(ia_p +l+i))-iShi(b(a_p+il+1)))}\Bigg],
    \end{aligned}
\end{equation}
where
$a_p=\frac{\omega_{cp}}{k_BT}$, $l=\omega_{cp}t$ and $b=\frac{\Delta_{\mathbf{k}}}{\omega_{cp}}$

\end{document}